# Gapless dispersive continuum in a modulated quantum kagome antiferromagnet

Asiri Thennakoon[1,6], Ryouga Yokokura[2,6], Yang Yang[1], Ryoichi Kajimoto[3], Mitsutaka Nakamura[3], Masahiro Hayashi[2], Chishiro Michioka[2], Gia-Wei Chern[1], Collin Broholm[4✉], Hiroaki Ueda[5✉], and Seung-Hun Lee[1✉]

[1]Department of Physics, University of Virginia; Charlottesville, Virginia, 22904, USA.

[2]Solid State Physics and Chemistry Lab, Division of Chemistry, Graduate School of Science, Kyoto University; Kyoto City 606-8502, Japan.

[3]Material and Life Science Division, J-PARC Center, Japan Atomic Energy Agency; Tokai, Ibaraki 319-1195, Japan.

[4]Institute for Quantum Matter and Department of Physics and Astronomy, The Johns Hopkins University; Baltimore, Maryland, 21218, USA.

[5]Institute for Advanced Materials Research and Development, Shimane University, Shimane 690-8504, Japan.

[6]These authors contributed equally: Asiri Thennakoon, Ryouga Yokokura

✉emails: broholm@jhu.edu; weda@mat.shimane-u.ac.jp; shlee@virginia.edu

**The pursuit of quantum spin liquid (QSL) states in condensed matter physics has drawn attention to kagome antiferromagnets (AFM) where a two-dimensional corner-sharing network of triangles frustrates conventional magnetic orders[1-4]. While quantum kagome AFMs based on $Cu^{2+}$ ($3d^9$, s=½) ions have been extensively studied[5,6], there is so far little work beyond copper-based systems. Here we present our bulk magnetization, specific heat and neutron scattering studies on single crystals of a new titanium fluorides $Cs_8RbK_3Ti_{12}F_{48}$ where $Ti^{3+}$ ($3d^1$, s = ½) ions form a modulated quantum kagome antiferromagnet that does not order magnetically down to 1.5 K. Our comprehensive map of the dynamic response function $S(\mathbf{Q},\hbar\omega)$ acquired at 1.5 K where the heat capacity is *T*-linear reveals a dispersive continuum emanating from soft lines that extend along (100). The data indicate fractionalized spinon-like excitations with quasi-one-dimensional dispersion within a quasi-two-dimensional spin system.**

Atomic magnetic moments in insulating solids generally develop static polarization below a temperature of order the inter-site interactions energy $J$. But in analogy with helium, which fails to solidify upon cooling, it may be possible to replace low temperature frozen magnetism by a intrinsically fluctuating state called a quantum spin liquid (QSL)[1-4]. Ever since this was proposed by Philip Anderson in 1973[7], numerous theoretical works have been reported to understand the nature of the QSL and determine whether it can be realized as a new state of matter. The theoretical consensus is that the spin liquid is a quantum mechanically long-ranged entangled state with excitations that are topological in nature. Two classes of quantum spin liquids have been proposed, depending on whether the excitation spectrum is gapped or gapless, which reflects different structures of the quantum entanglement[8-12]. Experimentally, quasi-one-dimensional materials have been identified that fail to order even for temperatures well below the interaction strength such as the uniform antiferromagnetic spin-1/2 chain[13,14]. However important features of spin-liquids including particle exchange processes are not possible in one dimension and long range order or a spin-Peierls lattice distortion eventually occurs for $T \ll J/k_B$. While the exactly solvable Kitaev model with frustrated bond dependent anisotropic interactions proves that a quantum spin liquid is possible in two dimensions, no materials realization has yet been identified and widely accepted. While an exact solution is not available, the quantum kagome Heisenberg antiferromagnet with isotropic antiferromagnetic near neighbor interactions on a two-dimensional corner-sharing network of triangles may also support a spin liquid and remains of great interest[15,16]. The most thoroughly studied kagome spin-liquid candidate so far is herbertsmithite, $ZnCu_3(OH)_6Cl_2$, where $Cu^{2+}$ ($d^9$, s = ½) ions provide the spin-1/2 degree of freedom[5,17]. Single-crystal neutron scattering studies have revealed a broad dispersionless continuum of spin excitations for wave vector transfer along lines that connect Γ points in neighboring Brillouin zones[15]. Beyond a low energy regime that is associated with inter-layer magnetic defects, and up to half of the magnetic bandwidth the dynamic correlation function appears to take a "local" factorized form $S(\mathbf{Q}, \hbar\omega) = f(\mathbf{Q})\ g(\hbar\omega)$ in herbertsmithite[2,18].

Recently several new kagome compounds $A_2B\mathrm{Ti_3F_{12}}$ were discovered where $Ti^{3+}$ $d^1$ forms the spin ½ and $F^-$ is the superexchange mediating ligand. These include inorganic compounds with $A$ = Rb, Cs and $B$ = Na, K, Li[19-21] and metal-organic compounds with $A$ = $CH_3NH_3$ and $B$ = Na, K[22,23]. These materials present new opportunities to explore kagome magnetism experimentally. In particular the periodic modulation of interactions within the kagome layers and the extreme two-dimensional character of the magnetism may promote spin-liquid-like characteristics. For while magnetization and specific heat studies report Curie-Weiss temperature, $\Theta_{CW}$, from $-30$ K to $-47$ K for the inorganic compounds[19-21], $\Theta_{CW} = -139.5$ K and $-83.5$ K for the metal-organic compounds $(CH_3NH_3)_2KTi_3F_{12}$ and $(CH_3NH_3)_2NaTi_3F_{12}$[22,23], respectively there are no indications of magnetic ordering. Various alternate ground states have been proposed ranging from a gapless disordered state ($Cs_2NaTi_3F_{12}$) and a gapped disordered state ($Cs_2KTi_3F_{12}$)[19], to a disordered state with weak spin-glass-like freezing ($(CH_3NH_3)_2B\mathrm{Ti_3F_{12}}$ ($B$ = Na, K))[23]. To go further with these promising QSL candidates, studies of their spin correlations at the atomic time and length scale are needed.

In this study we report a neutron scattering study of $Cs_8RbK_3Ti_{12}F_{48}$, which provides access to the wavevector ($\mathbf{Q}$) and energy ($\hbar\omega$) dependent dynamic spin correlation function of this new material in the titanium kagome family. Specifically, time-of-flight (TOF) neutron scattering measurements were performed on three co-mounted single crystals with a total mass of ~ 1 g. To map the entire band of magnetic excitations up to 15 meV with the appropriate energy resolutions, data with incident neutron energies, $E_i = (3, 4, 6, 10, 19)$ meV and corresponding energy resolutions $(0.11, 0.17, 0.27, 0.51, 1.20)$ meV were concurrently acquired. For an overview of the phase diagram, bulk, magnetization and specific heat capacity measurements were also performed on a single crystal. The experiments show that $Cs_8RbK_3Ti_{12}F_{48}$ does not develop magnetic order down to 3% of the Weiss temperature. The coexistence of a $T$-linear specific heat and linearly dispersive gapless continuum scattering emanating from lines in reciprocal space indicates a quasi-one-dimensional character though strong magnetic interactions define a two-dimensional network of interacting spins. These observations indicate that $Cs_8RbK_3Ti_{12}F_{48}$ may indeed form a quasi-two-dimensional gapless quantum spin liquid.

The crystal structure of $Cs_8RbK_3Ti_{12}F_{48}$ has the *R3m* space group with the hexagonal unit cell parameters of $a = 15.242(18)$ Å and $c = 18.477(19)$ Å (see Supplementary Information). The magnetic $Ti^{3+}$ ($s = ½$) ions form two-dimensional modulated kagome layers in the *ab*-plane (see the inset of Fig. 1a) that are well separated by nonmagnetic layers of alkali cations (Supplementary Fig. 2a). In the kagome layer, each $Ti^{3+}$ ion is surrounded by six $F^-$ ions, and the fluorine octahedra of neighboring $Ti^{3+}$ ions share one corner (see Supplementary Fig. 2b). The fluorine octahedra are compressed along the crystallographic *c*-axis so that the $d^1$ electron of the $Ti^{3+}$ ion occupies the $d_{xy}$ orbital that extends mainly in the *ab*-plane. The magnetic interactions between neighboring $Ti^{3+}$ ions arise from super-exchange interaction via their shared $F^-$ ligand. While the difference between the longest and the shortest bond length is just 0.47 %, the ∠Ti-F-Ti bond angles vary from $139.1^o$ to $148.1^o$ which Density Functional Theory projects to produce a variation in the super-exchange interactions (see Supplementary Information). The chemical unit cell in the *ab*-plane contains 2 x 2 unit cells of the simple uniform kagome lattice. The DFT results indicate a modulation of the exchange constants in the form of two inter-penetrating triangular superlattices; one superlattice shown as red solid and red dashed lines in the inset of Fig. 1a can be viewed as small triangles (formed by red dashed lines) separated by (10) and (01) from each other that are connected by red solid and red dashed lines, and the other superlattice shown as blue solid and blue dashed lines can be viewed similarly, i.e., small triangles (formed by blue solid lines). We denote wavevector transfer **Q** by Miller indices $(HK)_{Ti}$ in the corresponding reciprocal lattice. We also use Miller indices $\left(\frac{H}{2}\frac{K}{2}\right)_{s.kag}$ that refer to a reciprocal lattice with twice larger reciprocal lattice vectors $\left(a^*_{s.kag} = 2\ a^*_{Ti}\right)$. It is not only for the comparison with other kagome compounds but also because, as discussed later, despite the expected modulation of the super-exchange interactions the dynamic magnetic unit cell revealed by spin fluctuations is the same as that of the simple uniform kagome lattice.

Fig. 1a shows bulk susceptibility $\left(\chi = \frac{M}{H}\right)$ data obtained from a single crystal of $Cs_8RbK_3Ti_{12}F_{48}$ measured in a 1 tesla field. Fitting the high *T* data (inset) yields a Curie-Weiss temperature of $\Theta_{CW} = -47.4$ K which corresponds to an average near neighbor exchange constant of $\bar{J} = 4.1$ meV. The effective moment is $p_{eff} = 1.74\ \mu_B \approx g\sqrt{s(s+1)}\ \mu_B = \sqrt{3}\ \mu_B$ for $g = 2$ and $s = ½$ as expected for $Ti^{3+}$. $\chi$ does not exhibit sharp anomalies indicative of any magnetic phase transition down to 2 K. A broad maximum centered at around $T^* = 12$ K that is present for fields parallel and perpendicular to the *c*-axis is consistent with a thermal cross over from a paramagnetic to a QSL phase in a Heisenberg spin system. The blue line represents the theoretical susceptibility calculated using the exact diagonalization method for a 12-site uniform kagome cluster, which reproduces the susceptibility maximum with the nearest neighbor isotropic exchange constant of $J = 7.2$ meV[25,26]. Fig. 1b shows magnetic specific heat divided by *T*, $\frac{C_{mag}}{T}$, (red symbols) and the magnetic entropy, $\Delta S$, (blue symbols) as a function of *T*. Upon cooling, $\frac{C_{mag}}{T}$ begins to increase for *T* below 150 K and then increases more sharply below 50 K ≈ $|\Theta_{CW}|$. A rounded maximum near $T^* \sim 4.5$ K could be associated with entering a QSL regime. The change in magnetic entropy $\Delta S$ reaches a plateau value of $6.7\frac{Joule}{mol_{Ti}\ K}$ that is close to the expected full entropy value of $Rln(2s+1) \approx 5.76 \frac{Joule}{mol_{Ti}\ K}$. In the inset, $C_{mag}$ is plotted for low temperatures as a function of $T$, which shows a gradual increase with increasing $T$ up to 10 K, and no anomaly that might be associated with a phase transition. The black line in the inset is a fit to $C_{mag} = \gamma T^{\alpha}$ with $\gamma =$ $0.211(1)$ J/mole/$K^{1+\alpha}$ and $\alpha = 1.049(3)$ . Near proportionality of $C_{mag}$ with $T$ at low temperatures was also reported for $(CH_3NH_3)_2NaTi_3F_{12}$[22]. This contrasts with the quadratic behavior expected for two-dimensional linearly dispersive Goldstone modes of a rotational symmetry breaking ordered state. It is also distinct from herbersmithite where $C_{mag} \propto T^{1.5}$ at low temperatures[27]. $T$-linear magnetic specific heat is consistent with a finite density of fermionic quasi-particles at the chemical potential. It is a property of the Fermi liquid where the Sommerfeld constant of proportionality is a measure of the density of states at Fermi energy $E_F$. The Heisenberg AFM spin-1/2 chain with a Luttinger liquid ground state has a Sommerfeld constant $\gamma = \frac{2}{3} k_B R/J$ where $R$ is the gas constant, $k_B$ is the Boltzmann's constant and $J$ is the exchange constant[28,29]. Introducing the Curie Weiss estimate for $\bar{J} = 4.1$ meV yields $\gamma = 0.12$ J/mole/$K^2$ which remarkably shows this two-dimensional magnet actually has more low energy density of states than a one-dimensional spin system with the average near neighbor exchange constant. Thus, we should expect soft modes extending over areas in **Q**-space as for the spin-1/2 chain.

Fig. 2 shows the dynamic spin correlation function $S(\mathbf{Q}, \hbar\omega)$, obtained from magnetic neutron scattering data acquired at 1.5 K $< T^*$(Fig. 2a) and 20 K $> T^*$(Fig. 2b), both temperatures that are well below $|\Theta_{CW}|$. The horizontal momentum axis ($M_2 \rightarrow \Gamma_1 \rightarrow \Gamma_2 \rightarrow M_2 \rightarrow K_2$) traverses all high symmetry lines in the 2D Brillouin zone (the left inset of Fig. 2c) and the energy range extends from 0.1 meV to $4 \cdot k_B|\Theta_{CW}| \sim 16$ meV. At 1.5 K, $S(\mathbf{Q}, \hbar\omega)$ exhibits strong signals at low energies (Fig. 2a) that weaken considerably at 20 K (Fig. 2b) and therefore are identified as magnetic. Fig. 2a shows the magnetic excitation spectrum extends to ~ 10 meV ~ $2.5 k_B|\Theta_{CW}|$. Strong gapless spin fluctuations are apparent at $\Gamma_2 = (20)_{\mathrm{Ti}} = (10)_{\mathrm{s.kag}}$ which is the characteristic wavevector of the $q = 0$ structure that does not break translational symmetries. No such gapless fluctuations are found at $K_2 = \left(\frac{4}{3}\frac{4}{3}\right)_{\mathrm{Ti}} = \left(\frac{2}{3}\frac{2}{3}\right)_{\mathrm{s.kag}}$, which is the characteristic wavevector of the $\sqrt{3} \times \sqrt{3}$ structure. The characteristic slow $q = 0$ spin fluctuations indicated in the inset of Fig. 2c has all spins pointing in or out of the triangles that make up the kagome lattice. The fact that the strong gapless spin fluctuations are present at $\Gamma_2 = (20)_{\mathrm{Ti}} = (10)_{\mathrm{s.kag}}$ and they are absent $M_1 = (10)_{\mathrm{Ti}} = (\frac{1}{2}0)_{\mathrm{s.kag}}$ tells us that despite the expected modulation of the super-exchange interactions the dynamic magnetic unit cell is the same as that of the simple uniform kagome lattice and 2 x 2 smaller than that of the chemical unit cell. As shown in Supplementary Fig. 4, the spin wave theory with the DFT estimated $J$s yields strong gapless excitations at $M_1 =$ $(10)_{\mathrm{Ti}} = (\frac{1}{2}0)_{\mathrm{s.kag}}$ in addition to much more complex spin wave spectra than the experimental $S(\mathbf{Q}, \hbar\omega)$. This suggests that some aspects of the quantum kagome antiferromagnet survive bond modulation. And thus, we will later discuss a spin wave theory for the uniform $J$ to compare with the data.

Does $Cs_8RbK_3Ti_{12}F_{48}$ undergo a phase transition or develop any static magnetic correlations? To answer this, we plot along the same path as for the $\mathbf{Q} - \hbar\omega$ slices, the difference in the elastic scattering intensity, $S(\boldsymbol{Q})_{el} = \int_{-0.062meV}^{0.062meV} S(\boldsymbol{Q}, \hbar\omega) d(\hbar\omega)$ between $T = 1.5$ K and $T = 20$ K as black bullets in Fig.4a. Seeing no Bragg peaks in these data, places an upper limit of 0.05 $\mu_B$ on the moment size for the static magnetic order indicated in Fig. 2c. Furthermore, when averaging $S(\mathbf{Q},\hbar\omega)/|F(\mathbf{Q})|^2$ at 1.5 K over the Brillouin zone ($F(\mathbf{Q})$ is the magnetic form factor of $Ti^{3+}$) and integrating over energy, we find $\int_{0.15meV}^{13meV}\langle\frac{S(\boldsymbol{Q},\hbar\omega)}{F(\boldsymbol{Q})^2}\rangle\, d(\hbar\omega) = 0.49(2)/Ti^{3+}$ which exhausts the total moment sum rule: $\frac{2}{3}s(s+1) = 0.5/Ti^{3+}$. Thus, we conclude that $Cs_8RbK_3Ti_{12}F_{48}$ remains in a dynamic potentially spin-liquid state at least down to 1.5 K $< T^* \ll |\Theta_{CW}|$. Fig. 2a reveals a dispersionless mode centered at $\hbar\omega = \Delta \sim 2.4$ meV that extends along zone boundaries. Such a mode is reproduced within spin wave theory (white lines in Fig. 2) as the so-called weathervane mode (co-rotation of spins within **a**-oriented chains around the axis of the parallel spins that surround it, i.e., for instance in the inset of Fig. 2c the co-rotation of ABAB sublattices around the C spin) driven to finite energy transfer by the DM interactions (Fig. 2c inset).

The white lines in Fig. 2a and 2b are the dispersion relations obtained with a phenomenological spin wave model to be discussed in detail later. But while this model can be tuned to account for the dispersive bounds of the continuum, there is strong magnetic scattering between the dispersion curves where no spin waves are present in the model. Though the large unit cell allows for more modes than the model calculations, the absence of a static magnetic order makes conventional spin wave theory of limited use to describe magnetic excitations in $Cs_8RbK_3Ti_{12}F_{48}$.

Focusing on the $\mathbf{Q}$ −dependence of magnetic excitations we plot the dynamic spin correlation function $\int_{\hbar\omega_1}^{\hbar\omega_2} S(\mathbf{Q},\hbar\omega)d(\hbar\omega)$ in representative energy bands as color maps in $\mathbf{Q}$ space (Fig. 3) and as line cuts along high symmetry trajectories (Fig. 4). Dispersion is apparent in the evolution of $\mathbf{Q}$ dependence of dynamic correlations with $\hbar\omega$ that is apparent in these figures. At low energies and low $T = 1.5$ K, $S(\mathbf{Q},\hbar\omega \in [0.3,1.3]$ meV) (Fig. 3a, 4a) forms sharp ridge of intensity along the $\Gamma_2 \rightarrow M_2 \rightarrow \Gamma_2'$ trajectory, which indicate quasi-one-dimensional correlations extending along weathervane chains (insets in Fig. 2c). $S(\mathbf{Q},\hbar\omega \in [1.9,2.9]$ meV) (Fig. 3b, 4b) covers the dispersionless DM mode centered at $\hbar\omega\sim 2.4$ meV, which is strongest along $\Gamma_2 \rightarrow M_2 \rightarrow \Gamma_2'$ and exhibit a broad three-fold-like peak centered at the $K_2$ point. The 1D cut in Fig. 4b shows the intensity rises from $\Gamma_1$ in proportion to Q both at 1.5 K and at 20 K. $S(\mathbf{Q},\hbar\omega \in [6,7]$ meV), close to the top of the magnetic excitation bands, shows a broad ring around $\Gamma_2$ (Fig. 3c), which is not present at lower $\hbar\omega$ and indicates dispersive magnetic excitations. Dispersiveness in the high energy range is also apparent in the 1D cut of Fig. 4c which features two peaks surrounding $\Gamma_2$. It is interesting to note that the unit cell for the dynamic q = 0 structure is the same as that of the simple kagome and 2 x 2 smaller than that of the chemical unit cell of $Cs_8RbK_3Ti_{12}F_{48}$.

At 20 K, $S(\mathbf{Q},\hbar\omega)$ shows similar features for all energy ranges though much broader than for 1.5 K. For instance, $S(\mathbf{Q},\hbar\omega \in [0.3,1.3]$ meV) broadens asymmetrically toward $K_2$ rather than $K_1$ (Fig. 3d). $S(\mathbf{Q},\hbar\omega \in [1.9,2.9]$ meV) shows similar broadening that makes the three-fold-like peak more prominent around $K_2$ (Fig. 3e). In $S(\mathbf{Q},\hbar\omega \in [6,7]$ meV) the dispersive ring like features observed at 1.5 K are no longer visible in Figs. 3f and 4c. The broadening of all features in $S(\mathbf{Q},\hbar\omega)$ with increasing $\hbar\omega$ is evident throughout Fig. 4a-f.

Now let us turn to the $\hbar\omega$-dependence of the magnetic response function. Fig. 4d-e show $S(\mathbf{Q},\hbar\omega)$ for 1.5 K and 20 K at $\Gamma_2$ (Fig. 4d), $K_2$ (Fig. 4e), and $M_2$ (Fig. 4f). At 1.5 K (blue circles) $S(\mathbf{Q},\hbar\omega)$ exhibits gapless excitations at $\Gamma_2$ with little structure at the $\hbar\omega \approx \Delta$ weathervane DM mode. Intensity decreases steeply up to $\hbar\omega \sim 4$ meV and then weakens more slowly all the way up to $\hbar\omega \sim 13$ meV. The gradually decreasing scattering intensity for 4 meV $\lesssim \hbar\omega \lesssim 13$ meV at $\Gamma_2$ indicates the existence of a continuum of excitations because as shown in Fig. 2a there are no spin wave modes for $\hbar\omega > \Delta$ at $\Gamma_2$. At $K_2$ and $M_2$ (Fig. 4e and 4f), $S(\mathbf{Q}, \hbar\omega)$ exhibits a broad peak centered at $\hbar\omega = \Delta$ and a similar continuum for $4\ \text{meV} \lesssim \hbar\omega \lesssim 13$ meV as at $\Gamma_2$. The disappearance of the weathervane mode peak throughout the Brillouin zone upon heating (Fig. 4d-f) signals that its presence relies on correlations that are lost at 20 K.

To appreciate the exotic nature of dynamic spin correlations in $Cs_8RbK_3Ti_{12}F_{48}$, we first consider a phenomenological model of damped spin waves with short lifetime. Contrary to our experimental evidence, the LSW theory, of course, assumes the existence of a magnetic long-range order, and thus the theory is not expected to accurately reproduce the experimental data. Nevertheless, through comparison with the measured $\mathbf{Q}$-resolved excitation spectra, LSW theory allows us to estimate relevant interactions with greater specificity than can be obtained from susceptibility data. The minimal spin Hamiltonian used contains two terms, the isotropic nearest neighbor exchange coupling, $J$, and an antisymmetric Dzyaloshinskii-Moriya (DM) interaction, $\boldsymbol{D}_{ij}$,[30,31]

$$\mathcal{H} = J \sum_{<NN>} S_i \cdot S_j + \sum_{<NN>} \boldsymbol{D}_{ij} \cdot \left(S_i \times S_j\right). \qquad (1)$$

Another reason for us to have considered the simplest spin Hamiltonian is that the spin Hamiltonian with modulated $J$s determined by the DFT calculations yields a ground state and low energy gapless magnetic fluctuations at $\text{M}_1 = (10)_{\text{Ti}} = (\frac{1}{2}0)_{\text{s.kag}}$ as shown in Supplementary Fig. 4. The absence of the low energy gapless magnetic fluctuations at $\text{M}_1 = (10)_{\text{Ti}} = (\frac{1}{2}0)_{\text{s.kag}}$ in the experimental data (Fig. 2a) again indicates a remarkable robustness of the quantum kagome spin liquid state.

The white solid lines in Fig. 2a and 2b are the resulting spin wave dispersion relations for the $\mathbf{q} = 0$ state obtained with $J = 8.7$ meV and $\boldsymbol{D}_{ij} = D\,\hat{z}$ with $D = 0.23$ meV. The LSW dispersion relations catch some of the overall properties of the experimental $S(\mathbf{Q}, \hbar\omega)$, such as two dispersive modes, one gapless and the other gapped, coming out of the $\Gamma_2$ point and both dispersing all the way up to $\hbar\omega_{\boldsymbol{M},\text{top}} = 2Js\left(1 + \sqrt{3}\frac{D}{J}\right) \approx 9.1$ meV at $M$ and $\hbar\omega_{\boldsymbol{K},\text{top}} = Js\sqrt{\frac{3}{2}}\sqrt{3 + 6\frac{D^2}{J^2} + 5\sqrt{3}\frac{D}{J}} \approx 9.6$ meV at $K$ points, with a dispersionless mode centered at $\hbar\omega \sim \Delta = 3\sqrt{2}Js\sqrt{\frac{D}{J}\left(\frac{D}{J} + \frac{1}{\sqrt{3}}\right)} = 2.67$ meV over an extended area of $\mathbf{Q}$ space. The value of $J = 8.7$ meV exceeds the value of the average exchange constant obtained from the Curie-Weiss temperature, $\Theta_{CW} = -47.4$ K by a factor 2.1, which indicates a quantum renormalization of the magnon bandwidth. The analysis also identifies the dispersionless mode at $\Delta$ as the zero-energy weathervane mode of the kagome AFM lifted to a finite energy by the DM interaction. Several theoretical works on the quantum kagome antiferromagnet predict that magnetic long-range order will appear at low temperatures when the ratio $\frac{D}{J}$ exceeds a critical value that ranges from 0.012 (tensor-network study[32]) to 0.08 (Density Matrix Renormalization Group[33]) to 0.1 (exact diagonalization[34]) to 0.12 (functional renormalization group[35]). $\frac{D}{J} = 0.027$ obtained from our analysis for $Cs_8RbK_3Ti_{12}F_{48}$ is smaller than all but the tensor-network value, which supports the lack of spin freezing in this compound. A detailed description of the LSW calculations is in the Methods section.

We have also performed the Landau-Lifshitz dynamics (LLD) simulations to compute the finite-temperature neutron scattering intensity $S(\mathbf{Q}, \omega)$. An advantage of the damped LSW is that it provides an analytic dispersion relation for a given spin Hamiltonian. The advantage of LLD is that it does not require an ordered state and provides an estimate of temperature dependent damping. As shown in Extended data Fig. 4b-4d, the best fitting of the LLD simulations to the 1.5 K experimental data was obtained with an effective temperature of 2.2 K.

To what extent might damped spin waves be able to account for the observed magnetic excitations at 1.5 K? We calculated $S_{LSW}(\mathbf{Q}, \hbar\omega)$ for spin waves with and energy-dependent lifetime, $\tau = \frac{\hbar}{2\delta\epsilon_{\boldsymbol{k}}}$ where $\delta\epsilon_{\boldsymbol{k}}$ is the energy uncertainty that increases with the energy of the spin wave for wave vector $\boldsymbol{k}$, $\epsilon_{\boldsymbol{k}}$: $\delta\epsilon_{\boldsymbol{k}} = \eta \cdot \epsilon_{\boldsymbol{k}}$. The optimal coefficient $\eta = 0.4$ was determined for the resulting $S_{LSW}(\mathbf{Q}, \hbar\omega)$ to reproduce the broad peak around $\hbar\omega \sim \Delta$ at $K_2$ and $M$ points as shown in Fig. 4e and 4f. The corresponding optimal lifetime is $\tau \sim 0.4$ ps for $\epsilon_{\boldsymbol{k}} \approx 2$ meV and 0.09 ps for $\epsilon_{\boldsymbol{k}} \approx 9$ meV. For comparison, the long-lived spin waves are usually calculated with $\tau \gtrsim 1$ ps for all energies. Even though $S_{LSW}(\mathbf{Q}, \hbar\omega)$ captures some salient features of the experimental $S(\mathbf{Q}, \hbar\omega)$ as discussed before, the damped LSW theory cannot reproduce $S(\mathbf{Q}, \hbar\omega)$ well. Most notably, for $\hbar\omega < \Delta$, $S_{LSW}(\mathbf{Q}, \hbar\omega)$ exhibits sharp Goldstone modes around $\Gamma_2$ (see Fig. 3g), while $S(\mathbf{Q}, \hbar\omega)$ exhibits intersecting ridges of scattering extending along the $\Gamma_2 \rightarrow M \rightarrow \Gamma_2'$ directions (Fig. 3a). Furthermore, $S_{LSW}(\mathbf{Q}, \hbar\omega)$ cannot reproduce the continuum that is present in $S(\mathbf{Q}, \hbar\omega)$ for $\hbar\omega \gtrsim 5$ meV (see Fig. 4d-4f). As shown in Fig. 3j-3l and in Fig. 4, the LLD simulations at 2.2 K yield similar results to the damped LSW theory. However, neither theory fully reproduces the experimental data.

The failure of these phenomenological models indicates that spin waves are unstable and fractionalize into two spin-1/2 quasi-particles which is a hallmark of the quantum spin liquid. Let us compare our results with previous relevant experimental and theoretical works. The low energy $S(\mathbf{Q}, \hbar\omega)$ shown in Fig. 3a that exhibit strong signals at the $\Gamma_2$ points and along the straight lines connecting neighboring $\Gamma_2$ points is consistent with $S(\mathbf{Q}, \hbar\omega)$ reported for $ZnCu_3(OH)_6Cl_2$ (herbertsmithite)[5], though experimental challenges have so far precluded a detailed mapping of the full spectrum of magnetic excitations beyond the low energies as was possible here for $Cs_8RbK_3Ti_{12}F_{48}$. The similarity of the dynamic structure factors of these disparate materials in the low energy limit is indicative of generic features of quantum Heisenberg kagome antiferromagnets. The strong signals at the $\Gamma_2$ points have been reproduced and interpreted as spinon excitations by different theories of quantum spin liquids[31,36,37] (these papers use an extended BZ in which their $M_e$ points correspond to our $\Gamma_2$ points). Both theories[31,36], on the other hand, predicted strong signals along $\Gamma_2 \rightarrow K_2 \rightarrow \Gamma_2'$ in our BZ scheme yielding a triangular shaped strong signal around each $K_2$ point rather than along the straight lines connecting the $\Gamma_2$ points as found at low energies in both $ZnCu_3(OH)_6Cl_2$[5] and $Cs_8RbK_3Ti_{12}F_{48}$. As shown in Fig. 3b, for energies $\hbar\omega \sim \Delta$, $S(\mathbf{Q}, \hbar\omega)$ exhibits the theoretically predicted triangular shape signals around $K_2$ points which is also found for $ZnCu_3(OH)_6Cl_2$ when $S(\mathbf{Q}, \hbar\omega)$ is integrated over energy up to $\hbar\omega \sim J/2$.[5] The $S(\mathbf{Q}, \hbar\omega)$ reported for $ZnCu_3(OH)_6Cl_2$ for energies up to $\hbar\omega \sim J/2$ was interpreted as a dynamic magnetic response function of the "local" form $S(\mathbf{Q}, \hbar\omega) = f(\boldsymbol{Q})\ g(\hbar\omega)^2$. However, as shown in Extended Data Figs. 5 and 6, and discussed in Extended analysis and discussion, the magnetic excitations in $Cs_8RbK_3Ti_{12}F_{48}$ are not associated with decoupled dimers as suggested in ref. 5 or trimers. Instead, as shown in Fig. 3c (dispersion ring around the $\Gamma_1$ point for $\hbar\omega > J/2$), and in 4b and 4c (one-dimensional plots showing the dispersive peaks for $\Delta \lesssim \hbar\omega \lesssim J$), our measurement of $S(\mathbf{Q}, \hbar\omega)$ shows magnetic excitations in $Cs_8RbK_3Ti_{12}F_{48}$ are dispersive as theoretically predicted[38,39].

In conclusion, our experimental study of the new modulated quantum kagome antiferromagnet $Cs_8RbK_3Ti_{12}F_{48}$ shows there is no magnetic phase transition down to $T = 3\%|\Theta_{CW}|$ and yields an upper limit of 0.05 $\mu_B$ on $q = 0$ magnetic order. Our comprehensive measurement of the dynamic correlation function provides a first view of $S(\mathbf{Q}, \omega)$ for a kagome spin liquid candidate and shows the coexistence of a diffuse continuum with dispersive boundaries and a finite energy resonance covering most of the Brillouin zone save $\mathbf{Q} = 0$. Ridges of quasi-elastic scattering extending perpendicular to the **a**-axis indicate an effective one-dimensional character despite the fully two-dimensional nature of the modulated quantum kagome lattice. Comparison with a phenomenological damped linear spin wave theory provides a minimal spin Hamiltonian that consists of an isotropic nearest neighbor exchange coupling with $J = 8.7$ meV and a Dzyaloshinskii-Moriya interaction $\boldsymbol{D}_{ij} = D\,\hat{z}$ with $D = 0.23$ meV. The resulting value of $\frac{D}{J} = 0.027$ supports the lack of magnetic order[34,35,37]. Damped LSW theory however cannot reproduce the experimental excitation spectrum, especially the omnipresent continuum scattering, which indicates fractionalization. Continuum scattering within dispersive limits is qualitatively consistent with the theoretical two-spinon excitation spectra reported for U(1)[37,40] and $Z_2$ quantum spin liquids[38,39] (note again that their $M_e$ point corresponds to our $\Gamma_2$ point). The nearly *T*-linear specific heat at low temperatures is a consistent thermodynamic indication of the failure of the spin-wave picture that points to the existence of a spinon Fermi surface[41,42]. A quenched structural disorder may induce a magnetic glassy state in which specific heat would exhibit a *T*-linear behavior at low temperatures. On the other hand, such a glassy state would yield the typical FC-ZFC hysteresis in the bulk magnetization, which is inconsistent with our data. Our data thus establish $Cs_8RbK_3Ti_{12}F_{48}$ as a leading candidate for realization of a quantum spin liquid in a modulated kagome material.

# Methods

### Growth of $Cs_8RbK_3Ti_{12}F_{48}$ single crystals and crystal structure determination

Single crystals of $Cs_8RbK_3Ti_{12}F_{48}$ were grown using alkali metal chlorides as flux. For starting materials, alkali metal fluorides and chlorides were dried, and $TiF_3$ was purified before use. All the materials were mixed and loaded into a Ni crucible, which was then heated to 800°C and slowly cooled at a rate of 2°C/h under an Ar atmosphere. The flux was then removed using water to isolate the single crystals. The typical size of the resulting single crystals was $\approx 4 \times 4 \times 1$ mm$^3$ as shown in Supplementary Fig. 1. Three largest single crystals with a total mass of ~ 1 g were co-mounted for neutron scattering measurements.

Structural analysis of the single crystal was performed using a DIP3200 x-ray diffractometer (XRD) (Bruker AXS) with a Mo target. The structural parameters, including anisotropic displacement parameters, were refined using the full matrix least-square method as implemented in SHELXL-97. The observed and calculated nuclear structure factors, $|F_{obs}|$ and $|F_{calc}|$, respectively, are plotted against each other in Extended Data Fig. 1. According to the refinement, $Cs_8RbK_3Ti_{12}F_{48}$ has a space group of $R3m$ that lacks inversion symmetry. In Supplementary Table 1 and 2, all structural parameters such as the unit cell constants, the atomic coordinates, and the room temperature isotropic displacement parameters are provided using hexagonal notation. Supplementary Fig. 2a shows a polyhedral representation of the crystal structure where the Ti-F octahedra layers are separated from each other by Rb-Cs-F layers. Supplementary Fig. 2b highlights the kagome lattice formed by the Ti-F octahedra. In hexagonal notation, the kagome planes are the crystallographic *ab* plane.

### Density functional theory and magnetic exchange couplings

Based on the crystal structure determined by the x-ray refinement that is summarized in Supplementary Table 1 and 2, we performed density functional theory (DFT) calculations with OPENMX to estimate the exchange coupling strengths, $J$, for the four distinct near neighbor $Ti^{3+}$ pairs. The parametrization of Perdew and Zunger for the local density approximation (LDA) was chosen for the exchange-correlation functional.[43] The on-site Coulomb interactions were treated via a simplified DFT+$U$ formalism. The hoping parameter $t$ was obtained using the maximally localized Wannier orbital formalism.[44]

The exchange coupling constants $J$ for $Cs_2BTi_3F_{12}$ ($B$ = K, Na, Li) and $Rb_2NaTi_3F_{12}$ were calculated using a mapping method that requires quantum calculation of a supercell as reported in Ref. [20,21]. Since $Cs_8RbK_3Ti_{12}F_{48}$ has a chemical unit cell that is 2 x 2 larger than for those four systems, the mapping method is difficult to apply. Instead, we evaluated the exchange constant assuming that the coupling strength is mainly governed by the transfer integral $t$ for each bond. Firstly, we calculated the values of $t$ for $Cs_2BTi_3F_{12}$ ($B$ = K, Na, Li) and $Rb_2NaTi_3F_{12}$, and plotted their reported $J$ values obtained by the mapping method as a function of $t^2$. As shown in Supplementary Fig. 3a, the $J$ values show a good linear relation with $t^2$: $\frac{J}{K} = 6.64 \cdot \left(\frac{t}{meV}\right)^2 - 40.26$. We then calculated $t$ for the near neighbor $Ti^{3+}$-$Ti^{3+}$ spin pairs in $Cs_8RbK_3Ti_{12}F_{48}$, and obtained estimates for $J$ from this formula. Supplementary Fig. 3b shows the modulated kagome lattice of $Cs_8RbK_3Ti_{12}F_{48}$ with the estimated $J$. The red solid, red dashed, blue solid, and blue dashed lines represent the exchange interaction strengths of $J$ = (7.4, 6.3, 5.1, 3.9) meV, respectively. Note that there are triangles with uniform $J$s (represented by blue solid lines or by red dashed lines) and triangles with nonuniform $J$s (represented by two different types of lines). We would like to stress here that this simple DFT estimation of $J$s should be used only as a starting point for further detailed DFT studies of this compound.

### Bulk magnetization and specific heat measurements

Bulk magnetization and specific heat measurements were performed on a single crystal using a Quantum Design MPMS-XL system and a PPMS-14LHS system (Quantum Design), respectively, at the Research Center for Low Temperature and Materials Science, Kyoto University. Magnetic susceptibility measurement for the zero-field cooled (ZFC) condition was conducted in a heating process at the rate of 1.5 K/min below 60 K and at the rate of 3 K/min above 60 K. The field cooled (FC) data was taken in a cooling process at the same rates as those used for the ZFC data. The Curie-Weiss temperature was determined by fitting the bulk susceptibility data between 100 K and 300 K. The rise of $\chi(T)$ below 3.5 K is due to unavoidable imperfections in real crystals. The inferred molar abundance is approximately 1%. Heat capacity was measured at each temperature using a relaxation method with temperature rise of 2%. After that, the sample was heated to the next measuring temperature at the rate of 0.5-5 K/min, and the measurement was commenced when temperature became stable.

## Neutron scattering measurements

The Time-Of-Flight (TOF) neutron scattering experiment was conducted using the 4D-Space Access Neutron Spectrometer (4SEASONS) at the Japan Proton Accelerator Research Complex (J-PARC) Materials and Life Science Experimental Facility, Tokai, Japan[45]. The experiment was carried out on three single crystals that were co-aligned and attached using CYTOP (CTL-107M) to Al plates that in turn were attached to an Al sample holder. The total sample mass was $\sim 1g$. The Al sample holder was mounted in a $^4$He cryostat with a 1.5 K base temperature. The crystals were mounted with the crystallographic $c$ axis parallel with the incident neutron beam, which allowed us to integrate the intensity along the $c$ axis perpendicular to the kagome plane. At 4SEASONS, the neutron scattering data were collected simultaneously using five different incident neutron energies of $E_i =$ (3, 4, 6, 10, 19) meV.[46] These incident energies provided full width at half maximum energy resolutions of (0.11, 0.17, 0.27, 0.51, 1.20) meV, respectively, at the elastic line. This allowed us to map inelastic scattering over the entire band of magnetic excitations up to 15 meV with appropriate energy resolution throughout. Data were acquired at two temperatures, 1.5 K and 20 K, to examine spin fluctuations below and above the cross-over temperature $T^*$. At the end of the experiment, the empty cryostat without the crystals was measured for background subtraction using the same experimental setup. The data were analyzed using the Utsusemi software package[47].

## Absolute normalization of magnetic neutron scattering data

Normalization of the neutron scattering intensity data to absolute units for the scattering cross section can be accomplished through comparison to well-known standards including incoherent elastic scattering from vanadium, sample incoherent elastic scattering, sample elastic nuclear peaks, and sample phonon scattering[48]. We used the sample incoherent elastic scattering as described in Supplementary Information.

## Symmetrization of neutron scattering data

In neutron scattering experiment, we could measure $S(\mathbf{Q})$ over a limited area in the momentum space as shown in Extended Data Fig. 3a as an example. Utilizing that magnetic scattering in $Cs_8RbK_3Ti_{12}F_{48}$ has a six-fold symmetry, we constructed $S(\mathbf{Q})$ over the 360° angular range as shown in Extended Data Fig. 3b, which provides a comprehensive representation of $S(\mathbf{Q})$ across the entire angular range.

## Damped linear spin wave theory

Linear spin wave (LSW) analysis is done on a long-range ordered $q = 0$ 120° state with the minimal spin Hamiltonian,

$$\mathcal{H} = J \sum_{<NN>} s_i \cdot s_j + \sum_{<NN>} \boldsymbol{D}_{ij} \cdot (s_i \times s_j)$$

where $J = 8.7$ meV and $\boldsymbol{D}_{ij} = D\hat{z}$ with $D = 0.23$ meV. A $q = 0$ 120° state can be represented by a three-sublattice state within each unit cell

$$s_{i,A} = \left(-\frac{\sqrt{3}}{2}, -\frac{1}{2}, 0\right), \quad s_{i,B} = \left(\frac{\sqrt{3}}{2}, -\frac{1}{2}, 0\right), \quad s_{i,C} = (0,1,0).$$

Transformation to the spin fully polarized local coordinates $s_{i,A} = R_A\tilde{s}_{i,A}$, $s_{i,B} = R_B\tilde{s}_{i,B}$, $s_{i,C} = R_C\tilde{s}_{i,C}$ is accomplished with the following rotation matrices

$$R_A = \begin{pmatrix} \frac{1}{4} & -\frac{\sqrt{3}}{4} & \frac{\sqrt{3}}{2} \\ -\frac{\sqrt{3}}{4} & \frac{3}{4} & -\frac{1}{2} \\ \frac{\sqrt{3}}{2} & \frac{1}{2} & 0 \end{pmatrix}, \quad R_B = \begin{pmatrix} \frac{1}{4} & \frac{\sqrt{3}}{4} & \frac{\sqrt{3}}{2} \\ \frac{\sqrt{3}}{4} & \frac{3}{4} & -\frac{1}{2} \\ -\frac{\sqrt{3}}{2} & \frac{1}{2} & 0 \end{pmatrix}, \quad R_C = \begin{pmatrix} 1 & 0 & 0 \\ 0 & 0 & 1 \\ 0 & -1 & 0 \end{pmatrix}.$$

The Holstein-Primakoff (HP) transformation[51] rewrites spin operators in the fully polarized local coordinates in terms of HP bosons where we can expand $\tilde{s}_i^x$ and $\tilde{s}_i^y$ to linear order of $1/s$

$$\tilde{s}_{i,\alpha}^x \approx \sqrt{\frac{s}{2}}\left(a_{i,\alpha} + a_{i,\alpha}^\dagger\right), \quad \tilde{s}_{i,\alpha}^y \approx -i\sqrt{\frac{s}{2}}\left(a_{i,\alpha} - a_{i,\alpha}^\dagger\right), \quad \tilde{s}_{i,\alpha}^z = s - a_{i,\alpha}^\dagger a_{i,\alpha}.$$

After transforming the HF bosons in the momentum space, we obtain

$$\mathcal{H}_{LSW} = Js\sum_{\boldsymbol{k}} A_{\boldsymbol{k}}^\dagger M_{\boldsymbol{k}} A_{\boldsymbol{k}},$$

where $A_{\boldsymbol{k}}^\dagger = \left(a_{\boldsymbol{k},A}^\dagger, a_{\boldsymbol{k},B}^\dagger, a_{\boldsymbol{k},C}^\dagger, a_{-\boldsymbol{k},A}, a_{-\boldsymbol{k},B}, a_{-\boldsymbol{k},C}\right)$ and

$$M_{\boldsymbol{k}} = \begin{pmatrix} B_{\boldsymbol{k}} & C_{\boldsymbol{k}}^\dagger \\ C_{\boldsymbol{k}} & B_{-\boldsymbol{k}}^T \end{pmatrix}$$

with

$$B_{\boldsymbol{k}} = \begin{pmatrix} 1+\sqrt{3}\frac{D}{J} & -\frac{e^{i\frac{\pi}{3}}\left(1+e^{-i\,\boldsymbol{v}_1\cdot\boldsymbol{k}}\right)}{8}\left(1-\sqrt{3}\frac{D}{J}\right) & \frac{e^{i\frac{2\pi}{3}}\left(1+e^{-i\,\boldsymbol{v}_2\cdot\boldsymbol{k}}\right)}{8}\left(1-\sqrt{3}\frac{D}{J}\right) \\ -\frac{e^{-i\frac{\pi}{3}}\left(1+e^{-i\,\boldsymbol{v}_1\cdot\boldsymbol{k}}\right)}{8}\left(1-\sqrt{3}\frac{D}{J}\right) & 1+\sqrt{3}\frac{D}{J} & -\frac{e^{i\frac{\pi}{3}}\left(1+e^{-i\,\boldsymbol{v}_3\cdot\boldsymbol{k}}\right)}{8}\left(1-\sqrt{3}\frac{D}{J}\right) \\ \frac{e^{-i\frac{2\pi}{3}}\left(1+e^{-i\,\boldsymbol{v}_2\cdot\boldsymbol{k}}\right)}{8}\left(1-\sqrt{3}\frac{D}{J}\right) & -\frac{e^{-i\frac{\pi}{3}}\left(1+e^{i\,\boldsymbol{v}_3\cdot\boldsymbol{k}}\right)}{8}\left(1-\sqrt{3}\frac{D}{J}\right) & 1+\sqrt{3}\frac{D}{J} \end{pmatrix},$$

$$C_{\boldsymbol{k}} = \frac{\sqrt{3}}{8}\left(\sqrt{3}+\frac{D}{J}\right)\begin{pmatrix} 0 & -\left(1+e^{-i\,\boldsymbol{v}_1\cdot\boldsymbol{k}}\right) & e^{i\frac{\pi}{3}}\left(1+e^{-i\,\boldsymbol{v}_2\cdot\boldsymbol{k}}\right) \\ -\left(1+e^{i\,\boldsymbol{v}_1\cdot\boldsymbol{k}}\right) & 0 & e^{-i\frac{\pi}{3}}\left(1+e^{-i\,\boldsymbol{v}_3\cdot\boldsymbol{k}}\right) \\ e^{i\frac{\pi}{3}}\left(1+e^{i\,\boldsymbol{v}_2\cdot\boldsymbol{k}}\right) & e^{-i\frac{\pi}{3}}\left(1+e^{i\,\boldsymbol{v}_3\cdot\boldsymbol{k}}\right) & 0 \end{pmatrix}$$

and $\boldsymbol{v}_1 = (2,0,0)$, $\boldsymbol{v}_2 = \left(1,\sqrt{3},0\right)$, $\boldsymbol{v}_3 = \left(-1,\sqrt{3},0\right)$.

Using Bogoliubov transformation for bosons[52] to diagonalize the Hamiltonian at $\boldsymbol{k} = 0$, we obtain the energy of the dispersionless mode

$$\Delta = Js\left(3\sqrt{2}\sqrt{\frac{D}{J}\left(\frac{D}{J}+\frac{1}{\sqrt{3}}\right)}\right) \approx 2.67\ \text{meV},$$

and similarly, diagonalizations at $\boldsymbol{k} = (1,0)_{\text{Ti}}$ and $\left(\frac{4}{3},\frac{4}{3}\right)_{\text{Ti}}$ yield the maximum of two dispersive modes

$$\hbar\omega_{\boldsymbol{M},\text{top}} = 2Js\left(1+\sqrt{3}\frac{D}{J}\right) \approx 9.1\ meV,$$

$$\hbar\omega_{\boldsymbol{K},\text{top}} = Js\sqrt{\frac{3}{2}}\sqrt{3+6\frac{D^2}{J^2}+5\sqrt{3}\frac{D}{J}} \approx 9.6\ meV$$

along the $(H,0)$ and $(H,H)$ directions, respectively.

Next, we use the results from LSW analysis to compute the neutron scattering intensity. The magnetic neutron scattering intensity is computed by[53]

$$S(\boldsymbol{Q},\omega) \propto F(\boldsymbol{Q})^2 \sum_{\mu\nu} \left(\delta_{\mu\nu} - \frac{Q_\mu Q_\nu}{|\boldsymbol{Q}|^2}\right) s^{\mu\nu}(\boldsymbol{Q},\omega),$$

where $F(\mathbf{Q})$ denotes the form factor for $Ti^{3+}$, $Q_\mu, Q_\nu$ are the $\mu, \nu$ components of the $\mathbf{Q}$ vector, and most importantly, $s^{\mu\nu}(\mathbf{Q},\omega)$ is the dynamical structure factor

$$s^{\mu\nu}(\mathbf{Q},\omega) = \sum_{\alpha\beta} \int dt\ e^{-i\omega t}\ \langle s_\alpha^\mu(-\mathbf{Q},0) s_\beta^\nu(\mathbf{Q},t)\rangle,$$

which can be computed in terms of Bogoliubov bosonic quasiparticle correlations obtained from LSW analysis

$$\langle b_{\alpha,\boldsymbol{k}}^\dagger(t) b_{\beta,\boldsymbol{k}'}(0)\rangle = \delta_{\alpha\beta}\delta_{\boldsymbol{k}\boldsymbol{k}'} n(\omega_{\alpha,\boldsymbol{k}}) e^{-i\omega_{\alpha,\boldsymbol{k}} t},$$

$$\langle b_{\alpha,\boldsymbol{k}}(0) b_{\beta,\boldsymbol{k}'}^\dagger(t)\rangle = \delta_{\alpha\beta}\delta_{\boldsymbol{k}\boldsymbol{k}'} [n(\omega_{\alpha,\boldsymbol{k}}) + 1] e^{i\omega_{\alpha,\boldsymbol{k}} t},$$

with the Bose factor $n(\omega_{\alpha,\boldsymbol{k}}) = \left[e^{\hbar\omega_{\alpha,k}/(k_B T)} - 1\right]^{-1}$, where the momentum $\mathbf{k}$ is within the magnetic Brillouin zone such that $\mathbf{Q} - \mathbf{k}$ gives integer multiples of the reciprocal lattice vectors. At $T = 0$, we have $n(\omega_{\alpha,\boldsymbol{k}}) = 0$, so only the second term contributes to the dynamical structure factor. For generic $\boldsymbol{k}$ points, the LSW Hamiltonian is diagonalized numerically, so the transformations between spin operators and Bogoliubov quasiparticles are also obtained at each $\mathbf{k}$ point numerically.

To account for the finite lifetime of the single magnon excitation, we introduce a broadening factor (or energy uncertainty) proportional to the energy of the magnon mode $\delta\epsilon_{\alpha,\boldsymbol{k}} \propto \epsilon_{\alpha,\boldsymbol{k}}$ for a finite magnon lifetime $\tau_{\alpha,\boldsymbol{k}} = \frac{\hbar}{2\delta\epsilon_{\alpha,\boldsymbol{k}}}$ such that $\epsilon_{\alpha,\boldsymbol{k}}$ is replaced by $\epsilon_{\alpha,\boldsymbol{k}} + i\delta\epsilon_{\alpha,\boldsymbol{k}}$. This gives rise to the following form of the dynamic correlation function

$$S(\mathbf{Q},\omega) = \frac{F(\mathbf{Q})^2}{\pi} \sum_{\mu\nu,\alpha} \left(\delta_{\mu\nu} - \frac{Q_\mu Q_\nu}{|\mathbf{Q}|^2}\right) \left[L_{\boldsymbol{k}}^{\mu\nu}\right]_{\alpha+3,\alpha+3} \frac{\delta\epsilon_{\alpha,\boldsymbol{k}}}{\delta\epsilon_{\alpha,\boldsymbol{k}}{}^2 + \left(\hbar\omega - \epsilon_{\alpha,\boldsymbol{k}}\right)^2},$$

where $L_{\boldsymbol{k}}^{\mu\nu}$ denotes the numerical transformation between the spin operators and the Bogoliubov bosonic quasiparticle operators (the subscript $\alpha + 3$ indicates the summation is only over the lower half diagonal elements at $T = 0$).

**Calculation for the finite-temperature neutron response**

The finite-temperature neutron scattering intensity $S(\mathbf{Q},\omega)$ can be computed through simulating the system using Landau-Lifshitz dynamics (LLD)

$$\frac{ds_i}{dt} = -s_i \times B_i$$

where the effective field is $B_i = -d\mathcal{H}/ds_i$ and $\mathcal{H}$ is the spin Hamiltonian. For a given initial state $\{s_i(t=0)\}$, the above LL equation can be integrated to produce "trajectories" of spins $\{s_i(t)\}$. The information of excited states in the initial state is encoded in these trajectories. To this end, we write the lattice site index as $i = (\boldsymbol{R}, \alpha)$, where $\boldsymbol{R}$ is the position vector of a given unit cell and $\alpha$ is the sublattice index. The Fourier transform of the spin trajectories are given by

$$s_\alpha(\boldsymbol{Q}, t) = \frac{1}{\sqrt{N}} \sum_i s_\alpha(\boldsymbol{R}, t)\, e^{-i\boldsymbol{Q}\cdot\boldsymbol{R}},$$

The dynamical structure factor is then given by

$$s^{\mu\nu}(\mathbf{Q}, \omega) = \sum_{\alpha\beta} \int dt\ e^{-i\omega t}\ \langle s_\alpha^\mu(-\mathbf{Q}, 0) s_\beta^\nu(\mathbf{Q}, t)\rangle,$$

Importantly, here the expectation value $\langle \cdots \rangle$ is taken over the different initial conditions which are sampled either using Monte Carlo simulations or stochastic Landau-Lifshitz-Gilbert simulations at a given temperature $T$. It is worth noting that the system energy is conserved in LLD. The dependence on temperature, which in turn determines the degree of spin fluctuations, solely comes from the sampled initial states. Moreover, since the sampled initial spins could be in a classical liquid phase, the LLD approach does not require the assumption of a long-range magnetic order. As a result, LLD methods have been employed in the calculation of $s^{\mu\nu}(\mathbf{Q}, \omega)$ for classical spin liquids such as in frustrated magnets. Using the LLD methods provided in the SUNNY package (https://github.com/SunnySuite/Sunny.jl), we computed the intensity of magnetic inelastic neutron scattering $S(\mathbf{Q}, \omega)$ from 15 independent simulation runs with $N = 100 \times 100 \times 3$ spins at $T =$ 1.5 K, 2.2 K, and 3.0 K.

## Extended analysis and discussion

### Dimer and trimer models

In neutron scattering work on herbertsmithite, $ZnCu_3(OH)_6Cl_2$,[5] the energy integrated magnetic response function was successfully compared to a noninteracting dimer model. The associated non-dispersive "local" form of the response function[2] might be associated with a disordered random singlet state or spinon-vison interactions[36]. In our experiments on $Cs_8RbK_3Ti_{12}F_{48}$ it has been possible to cover the entire range of the magnetic bandwidth for a kagome material, which may allow distinguishing such scenarios.

Here we examine whether our neutron scattering data for $Cs_8RbK_3Ti_{12}F_{48}$ are consistent with local excitations as for a random singlet phase. Two local excitations were considered, dimers and 120° trimers. Firstly, we found that the models of dimers and trimers yield the same magnetic structure factor, i.e., they are not distinguishable through neutron scattering. Secondly, the energy integrated equal time magnetic response function measured for $Cs_8RbK_3Ti_{12}F_{48}$ cannot be reproduced by the local dimer/trimer model. Though damped linear spin wave theory cannot account for the broad continuum scattering, it describes the equal-time response function better than local models of dimers and trimers.

As shown in Supplementary Fig. 6a, considering their phase factors there are two types of 120° trimers where three spins form a 120° spin configuration as shown in the inset. The magnetic structure factor of the first trimer circled by the red dashed line is

$$\boldsymbol{F}_{trimer,1}(h,k) = |\boldsymbol{A}|e^{i\pi k}\hat{y} + |\boldsymbol{B}|e^{i\frac{\pi}{2}(h+2k)}\left(-\frac{\sqrt{3}}{2}\hat{x} - \frac{1}{2}\hat{y}\right) + |\boldsymbol{C}|e^{i\frac{\pi}{2}k}\left(\frac{\sqrt{3}}{2}\hat{x} - \frac{1}{2}\hat{y}\right).$$

Or assuming that $|\boldsymbol{A}| = |\boldsymbol{B}| = |\boldsymbol{C}| = 1$,

$$\boldsymbol{F}_{trimer,1}(h,k) = e^{i\pi k}\left[\left(-\frac{\sqrt{3}}{2}e^{i\frac{\pi}{2}h} + \frac{\sqrt{3}}{2}e^{-i\frac{\pi}{2}k}\right)\hat{x} + \left(1 - \frac{1}{2}e^{i\frac{\pi}{2}h} - \frac{1}{2}e^{-i\frac{\pi}{2}k}\right)\hat{y}\right].$$

Similarly, for the second trimer circled by the blue dashed line,

$$\boldsymbol{F}_{trimer,2}(h,k) = e^{i\pi h}\left[\left(-\frac{\sqrt{3}}{2}e^{-i\frac{\pi}{2}h} + \frac{\sqrt{3}}{2}e^{i\frac{\pi}{2}k}\right)\hat{x} + \left(1 - \frac{1}{2}e^{-i\frac{\pi}{2}h} - \frac{1}{2}e^{i\frac{\pi}{2}k}\right)\hat{y}\right].$$

Thus, the neutron scattering intensity from the two noninteracting trimers becomes

$$I_{trimer}(h,k) \propto \left|\boldsymbol{F}_{trimer,1}(h,k)\right|^2 + \left|\boldsymbol{F}_{trimer,2}(h,k)\right|^2 \propto \left(3 - cos\frac{\pi}{2}h - cos\frac{\pi}{2}k - cos\frac{\pi}{2}(h+k)\right).$$

Supplementary Fig. 6b, on the other hand, shows three types of decoupled dimers. Since the structure factor squared of a dimer is $|\boldsymbol{F}_{dimer}(\boldsymbol{Q})|^2 \propto (1 - \cos(\boldsymbol{Q}.\boldsymbol{r}))$[49,50], the neutron scattering intensity from the three noninteracting dimers becomes

$$I_{dimer}(h,k) \propto \left|\boldsymbol{F}_{dimer,1}(h,k)\right|^2 + \left|\boldsymbol{F}_{dimer,2}(h,k)\right|^2 + \left|\boldsymbol{F}_{dimer,3}(h,k)\right|^2 \propto \left(1 - cos\frac{\pi}{2}(h+k)\right) + \left(1 - cos\frac{\pi}{2}k\right) + \left(1 - cos\frac{\pi}{2}h\right) = \left(3 - cos\frac{\pi}{2}h - cos\frac{\pi}{2}k - cos\frac{\pi}{2}(h+k)\right).$$

Therefore, $I_{trimer}(h,k)$ and $I_{dimer}(h,k)$ are exactly the same and cannot be distinguished by neutron scattering.

**Q-dependence of the equal-time response function and propagating excitation**

Extended Data Fig. 5a shows a contour map of the experimental equal-time response function $S(\mathbf{Q}) = \int_{0.3meV}^{13meV} S(\mathbf{Q},\hbar\omega)\, d(\hbar\omega)$ at 1.5 K that provides us with important information on the putative quantum spin liquid state of $Cs_8RbK_3Ti_{12}F_{48}$. $S(\mathbf{Q})$ exhibits three salient features: A global maximum at $\Gamma_2$, a strong signal along the $\Gamma_2 \rightarrow M_2 \rightarrow \Gamma_2'$ direction, and a broad bump near $K_2$. Extended Data Fig. 5b and 5c show $S(\mathbf{Q})_{LSW}$ for damped LSWs and the structure factor for decoupled dimers/trimers, $S(\mathbf{Q})_{dimer/trimer}$, respectively. $S(\boldsymbol{Q})_{dimer/trimer}$ displays a three-fold symmetric peak centered at $K_2$ and a weaker signal at around $\Gamma_2$, which is inconsistent with the experimental $S(\boldsymbol{Q})$. Consistent with the data, $S(\mathbf{Q})_{LSW}$ for the dampled spin-wave model, on the other hand, shows a strong signal at $\Gamma_2$.

For a more detailed comparison of these models with the experimental data, Extended Data Fig. 6 shows $S(\mathbf{Q})$, $S(\mathbf{Q})_{LSW}$, and $S(\mathbf{Q})_{dimer/trimer}$ along the $\mathrm{M}_2 \rightarrow \Gamma_1 \rightarrow \Gamma_2 \rightarrow \mathrm{M}_2 \rightarrow \mathrm{K}_2$ path in the 2D Brillouin zone. $S(\mathbf{Q})$ is strong at $\Gamma_2$ and along $\Gamma_2 \rightarrow \mathrm{M}_2$. It is stronger at $\mathrm{K}_2$ than at $\mathrm{K}_1$. $S(\mathbf{Q})_{LSW}$ shows strong maxima at $\Gamma_2$ points, while it is sharper than $S(\boldsymbol{Q})$ along all directions except $\mathrm{M}_2 \rightarrow \mathrm{K}_2$. For $\mathrm{M}_2 \rightarrow \mathrm{K}_2$, $S(\mathbf{Q})$ is stronger at $\mathrm{M}_2$ than at $\mathrm{K}_2$, while $S(\mathbf{Q})_{LSW}$ is stronger at $\mathrm{K}_2$ than at $\mathrm{M}_2$. $S(\mathbf{Q})_{dimer/trimer}$ reproduces the broad features along $\mathrm{M}_2 \rightarrow \Gamma_1 \rightarrow \Gamma_2$. However, $S(\mathbf{Q})_{dimer/trimer}$ does not reproduces the sharp peak of $S(\mathbf{Q})$ centered at $\Gamma_2$, and contrary to the experimental data for $(\boldsymbol{Q})$ yields no modulation along $\Gamma_2 \rightarrow \mathrm{M}_2$ beyond the $Ti^{3+}$ form factor. Thus, we conclude that magnetic excitations in $Cs_8RbK_3Ti_{12}F_{48}$ are not associated with decoupled dimers or trimers.

## Acknowledgments

We thank J. Wen, Y. Lee, and C. Batista for discussion. A.T. and S.H.L. were supported by the U.S. Department of Energy, Office of Science, Basic Energy Sciences, Basic Energy Sciences, through DE-SC-001614. R.Y., M.H., C.M, and H.U. were supported by JSPS KAKENHI Grant Number JP23K25820. Y.Y. and G.-W.C. were partially supported by the US Department of Energy, Basic Energy Sciences under Contract No. DE-SC0020330. C.B. was supported by the Gordon and Betty Moore Foundation EPIQS program under GBMF9456. The neutron scattering experiments at the Material and Life Science Experimental Facility, Japan Proton Accelerator Research Complex was performed under a user program (Proposal No. 2023BU0101).

## Author contributions

H.U. and S.H.L. initiated and designed the research. All authors contributed to carry out the experiments. A.T., Y.Y., G.W.C., C.B, and S.H.L. analyzed neutron scattering data. All authors contributed to the writing of the manuscript.

## Competing interests

The authors declare no competing interests.

## Additional Information

### Supplementary Information

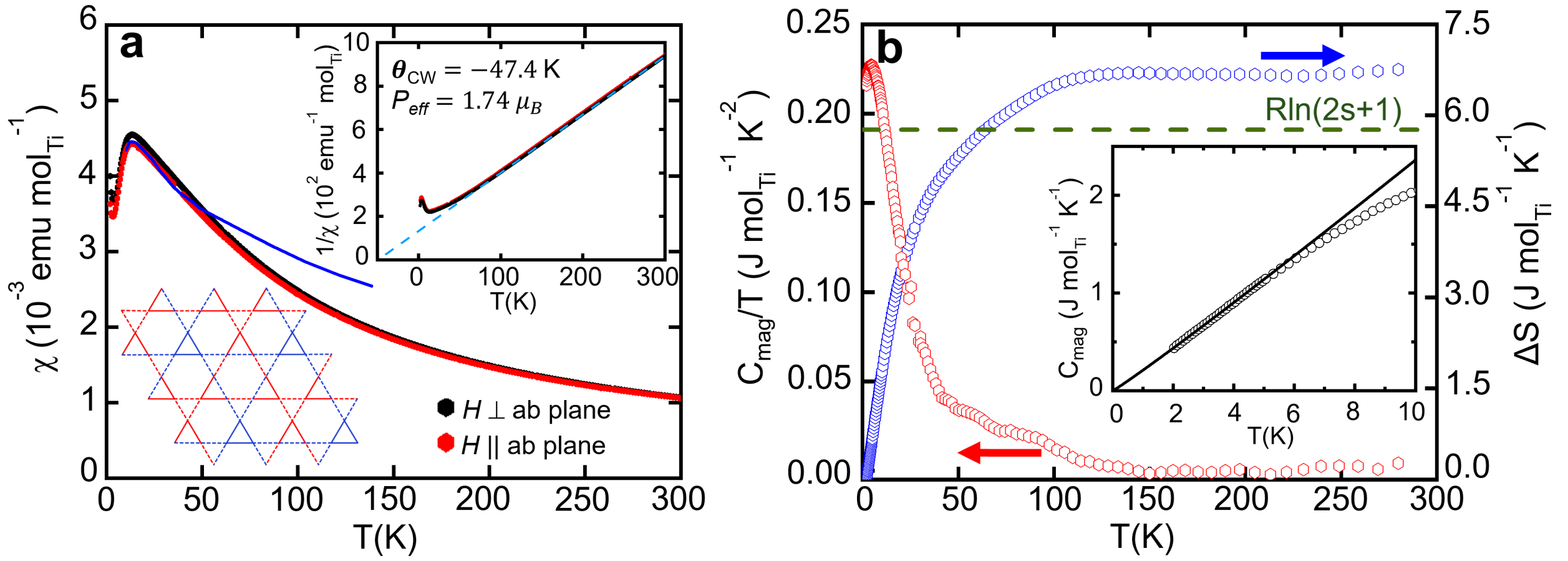

**Fig. 1 | Bulk magnetic susceptibility and specific heat. a**, Bulk susceptibility, $\chi = M/H$, as a function of temperature ($T$), for a single crystal of $Cs_8RbK_3Ti_{12}F_{48}$. The left bottom insert shows the kagome lattices formed by $Ti^{3+}$ ions. The crystal structure of $Cs_8RbK_3Ti_{12}F_{48}$ is obtained by x-ray refinement and is described in the Supplementary Information. Near neighbor exchange couplings predicted by Density Functional Theory are drawn as red solid, red dashed, blue solid, and blue dashed lines for the coupling strengths of $J = (7.4, 6.3, 5.1, 3.9)$ meV, respectively (see the Supplementary Fig. 3b). Black line and red line are data taken under an external magnetic field of $\mu_0 H = 1$ T perpendicular to and parallel to the *ab*-plane, respectively. The blue line represents the theoretical susceptibility calculated by the exact diagonalization method for a 12-site kagome cluster with an exchange constant of $J = 7.2$ meV[25,26]. The inset on the top right shows $1/\chi$ vs $T$ and the dashed blue line is a fit to the Curie-Weiss law, with Curie-Weiss temperature $\Theta_{CW} = -47.4$ K and an effective moment of $P_{eff} = 1.74\,\mu_B \approx g\sqrt{s(s+1)}\mu_B$ with the $g$-factor $g = 2$ and $s = 1/2$. **b**, Magnetic specific heat divided by $T$, $C_{mag}/T$ (red symbols), and spin entropy $\Delta S$ (blue symbols) as a function of temperature. The phonon contribution to the specific heat capacity was measured for nonmagnetic isostructural $Cs_2KGa_3F_{12}$ and subtracted to obtain $C_{mag}$. The horizontal dashed line is the total entropy, $S_{tot} = R\ln(2s+1)$, expected for $s = 1/2$. The insert shows $C_{mag}$ at low temperatures, and the black line is a fit to $C_{mag} = \gamma T^{\alpha}$ with $\gamma = 0.211(1)$ J/mole/$K^{1+a}$ and $\alpha = 1.049(3)$.

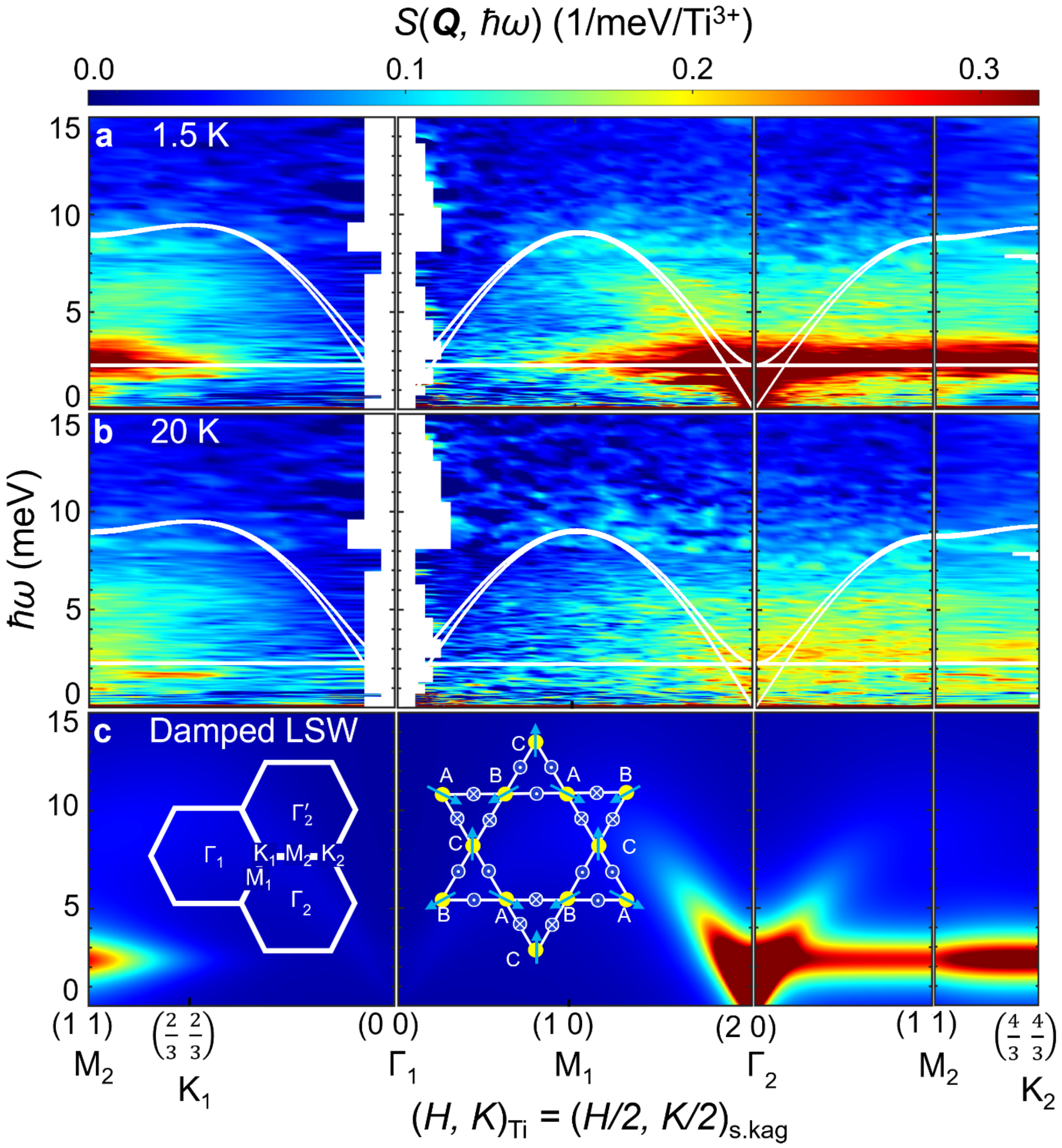

**Fig. 2 | Dynamic spin correlation function for $Cs_8RbK_3Ti_{12}F_{48}$.** Color contour maps of the normalized dynamic spin correlation function, $S(\mathbf{Q}, \hbar\omega)$, versus momentum ($\mathbf{Q}$) and energy transfer ($\hbar\omega$) along the $M_2 \to \Gamma_1 \to \Gamma_2 \to M_2 \to K_2$ path in the 2D Brillouin zone at **a**, 1.5 K and **b**, 20 K. In those panels, the white lines represent the dispersion relations of linear spin waves (LSW) predicted for the spin Hamiltonian of Eq. (1) with $J = 8.7$ meV and $\boldsymbol{D}_{ij} = D\,\hat{z}$ with $D = 0.23$ meV (see main text). **c**, Contour maps of $S_{LSW}(\mathbf{Q}, \hbar\omega)$, calculated for this model using spin wave theory and adding a phenomenological energy dependent damping term with lifetime $\tau = \frac{\hbar}{2\delta\epsilon_{\boldsymbol{k}}}$ where $\delta\epsilon_{\boldsymbol{k}} = 0.4 \cdot \epsilon_{\boldsymbol{k}}$. See main text for details. The insets show the Brillouin zone boundaries

of the simple kagome lattice and a $q = 0$ magnetic structure. The symbols $\odot, \otimes$ represent the direction of the DM vectors used in the minimal model.

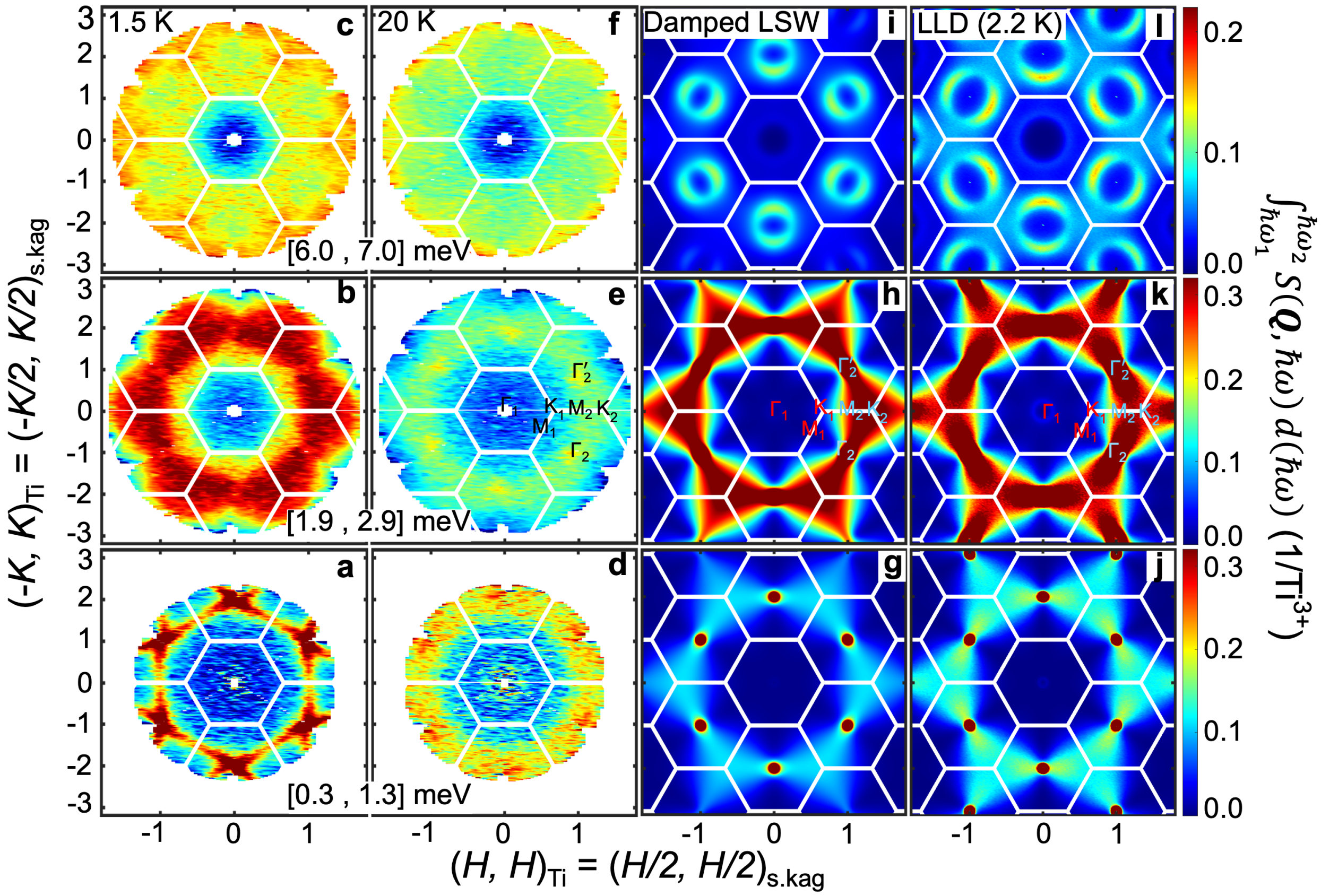

**Fig. 3 | Q-dependences of dynamic spin correlations in $Cs_8RbK_3Ti_{12}F_{48}$ for different energy ranges.** Symmetrized color contour maps of the $\hbar\omega$ -integrated response function $\int_{\hbar\omega_1}^{\hbar\omega_2} S(\boldsymbol{Q},\hbar\omega)\, d(\hbar\omega)$, **a-c**, for 1.5 K and **d-f**, for 20 K, and **g-i**, those of the damped linear spinwave theory with lifetime $\tau = \frac{\hbar}{2\delta\epsilon_{\boldsymbol{k}}}$ where $\delta\epsilon_{\boldsymbol{k}} = 0.4 \cdot \epsilon_{\boldsymbol{k}}$, **j-l**, those of Landau-Lifshitz dynamics (LLD) simulations at 2.2 K, obtained for three integration ranges: **a, d, g, j**, $[\hbar\omega_1, \hbar\omega_2] = [0.3, 1.3]$ meV; **b, e, h, k**, $[\hbar\omega_1, \hbar\omega_2] = [1.9, 2.9]$ meV; and **c, f, i, l**, $[\hbar\omega_1, \hbar\omega_2] = [6.0, 7.0]$ meV.

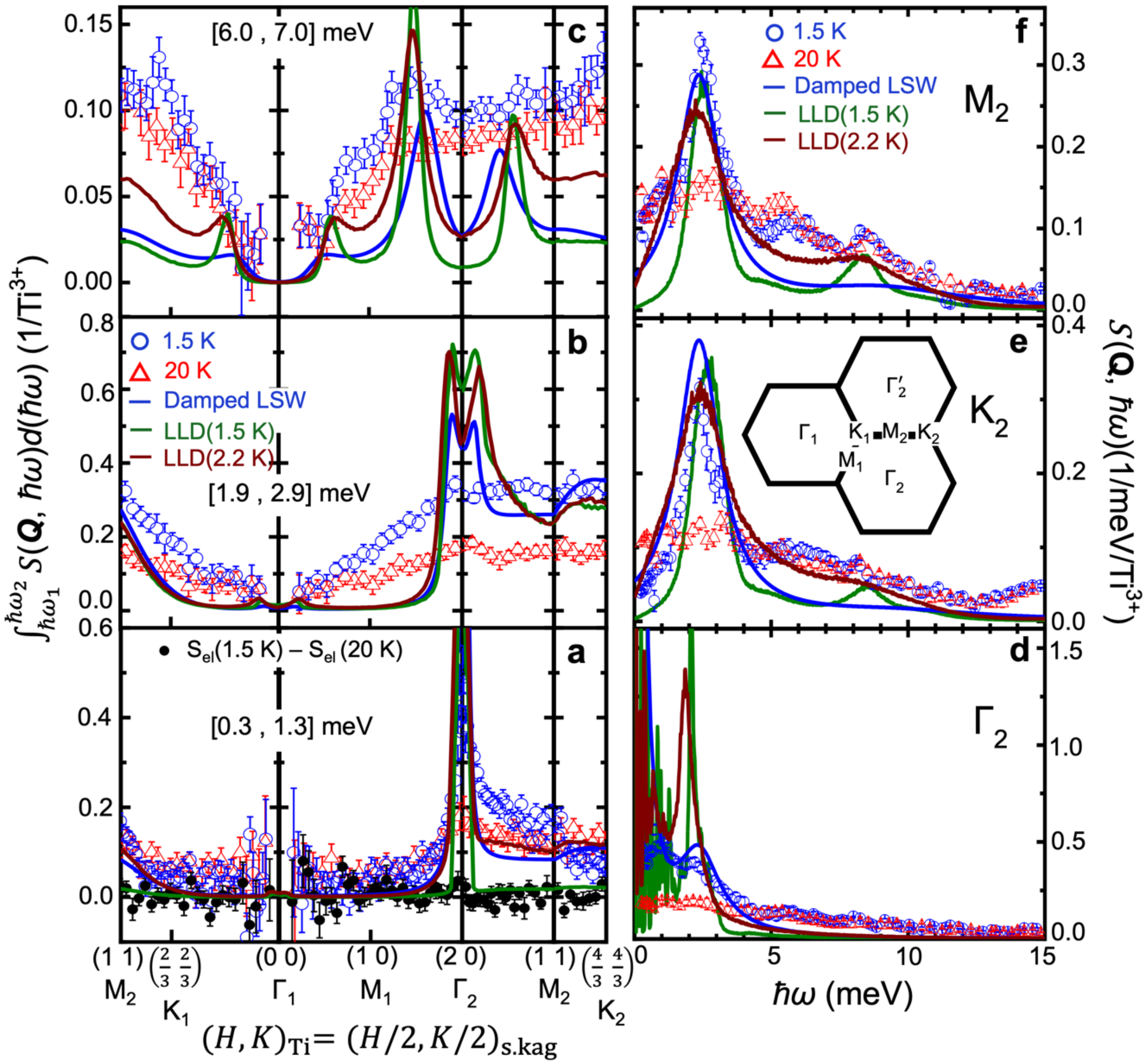

**Fig. 4 | Q- and $\hbar\omega$- dependence of the dynamic spin correlations in $Cs_8RbK_3Ti_{12}F_{48}$. a-c**, **Q**-dependence of the $\hbar\omega$-integrated response function $\int_{\hbar\omega_1}^{\hbar\omega_2} S(\boldsymbol{Q},\hbar\omega)\,d(\hbar\omega)$ along $\mathrm{M}_2 \to \Gamma_1 \to \Gamma_2 \to \mathrm{M}_2 \to \mathrm{K}_2$, with **a**, $[\hbar\omega_1, \hbar\omega_2] = [0.3,1.3]$ meV; **b**, [1.9,2.9] meV; and **c**, [6.0,7.0] meV. **d-f**, $\hbar\omega$-dependence of $S(\mathbf{Q},\hbar\omega)$ at three different high symmetry $\boldsymbol{Q}$ points: **d**, $\Gamma_1$; **e**, $K_2$; and **f**, M. Here to increase the statistics, we integrated $S(\mathbf{Q},\hbar\omega)$ over $(h \pm 0.1, k \pm 0.1)_{\mathrm{Ti}}$ for $\mathbf{Q} = (h,k)_{\mathrm{Ti}}$. In all panels, blue circles and red triangles are data for 1.5 K and 20 K, respectively. Black filled circles in **a** are the difference in the elastic signal between 1.5 K and 20 K, $S_{el}(\mathbf{Q}, 1.5\ \mathrm{K}) - S_{el}(\mathbf{Q}, 20\ \mathrm{K})$, where $S_{el}(\mathbf{Q}) = \int_{-0.062meV}^{0.062meV} S(\mathbf{Q},\hbar\omega)\,d(\hbar\omega)$, showing there is no spin freezing, either short-ranged or long-ranged, along the high symmetry directions of $\mathrm{M}_2 \to \Gamma_1 \to \Gamma_2 \to \mathrm{M}_2 \to \mathrm{K}_2$. Blue solid lines represent the corresponding response function calculated for the damped LSWs with lifetime $\tau = \frac{\hbar}{2\delta\epsilon_{\boldsymbol{k}}}$ where $\delta\epsilon_{\boldsymbol{k}} = \eta\epsilon_{\boldsymbol{k}}$ with $\eta = 0.4$, while green and deep violet lines represent the results of LLD simulations at 1.5 K and 2.2 K, respectively.

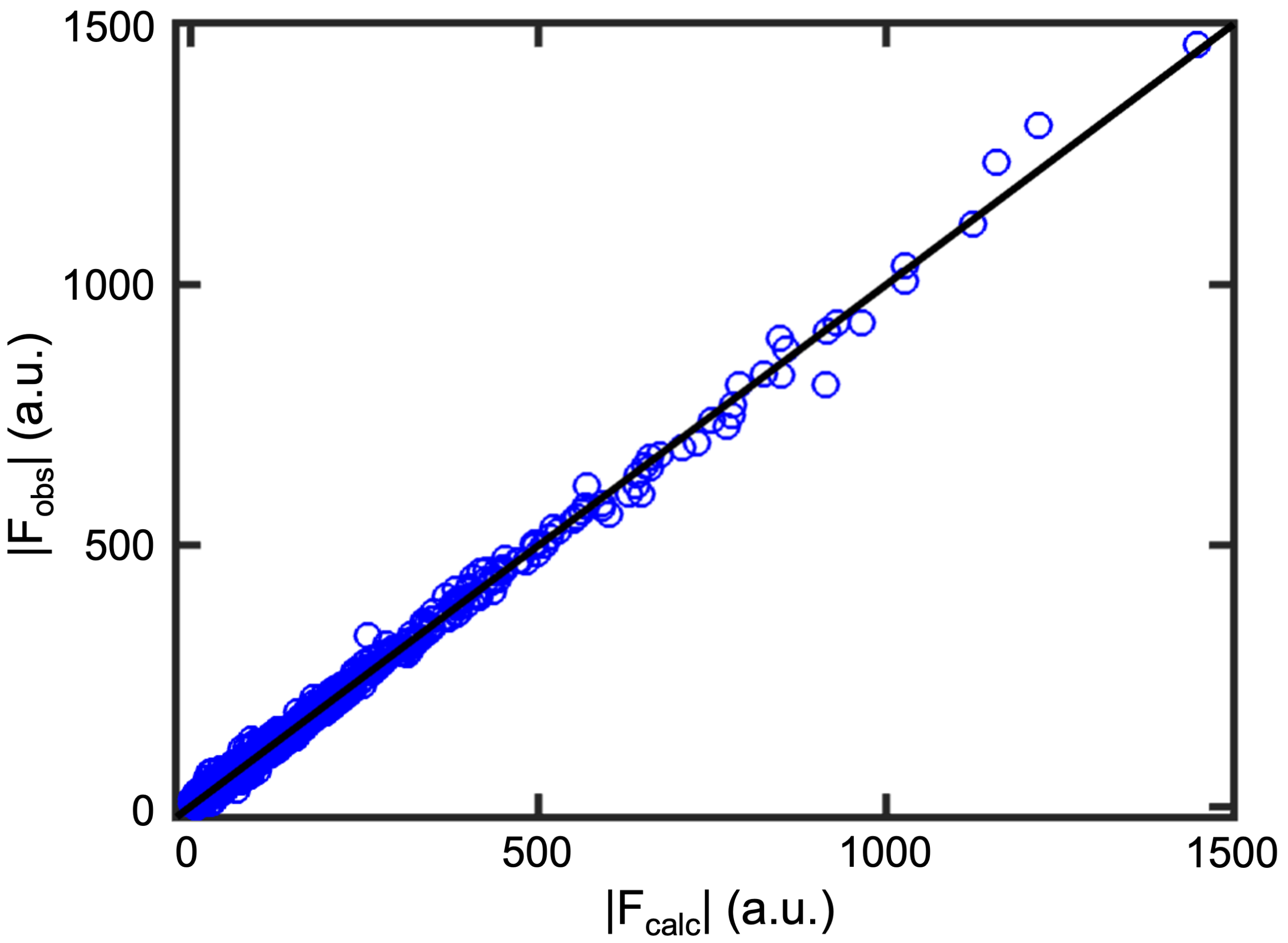

**Extended Data Fig. 1 | Single crystal x-ray data.** The observed and calculated nuclear structure factors, $|F_{obs}|$ and $|F_{calc}|$, respectively, are plotted against each other.

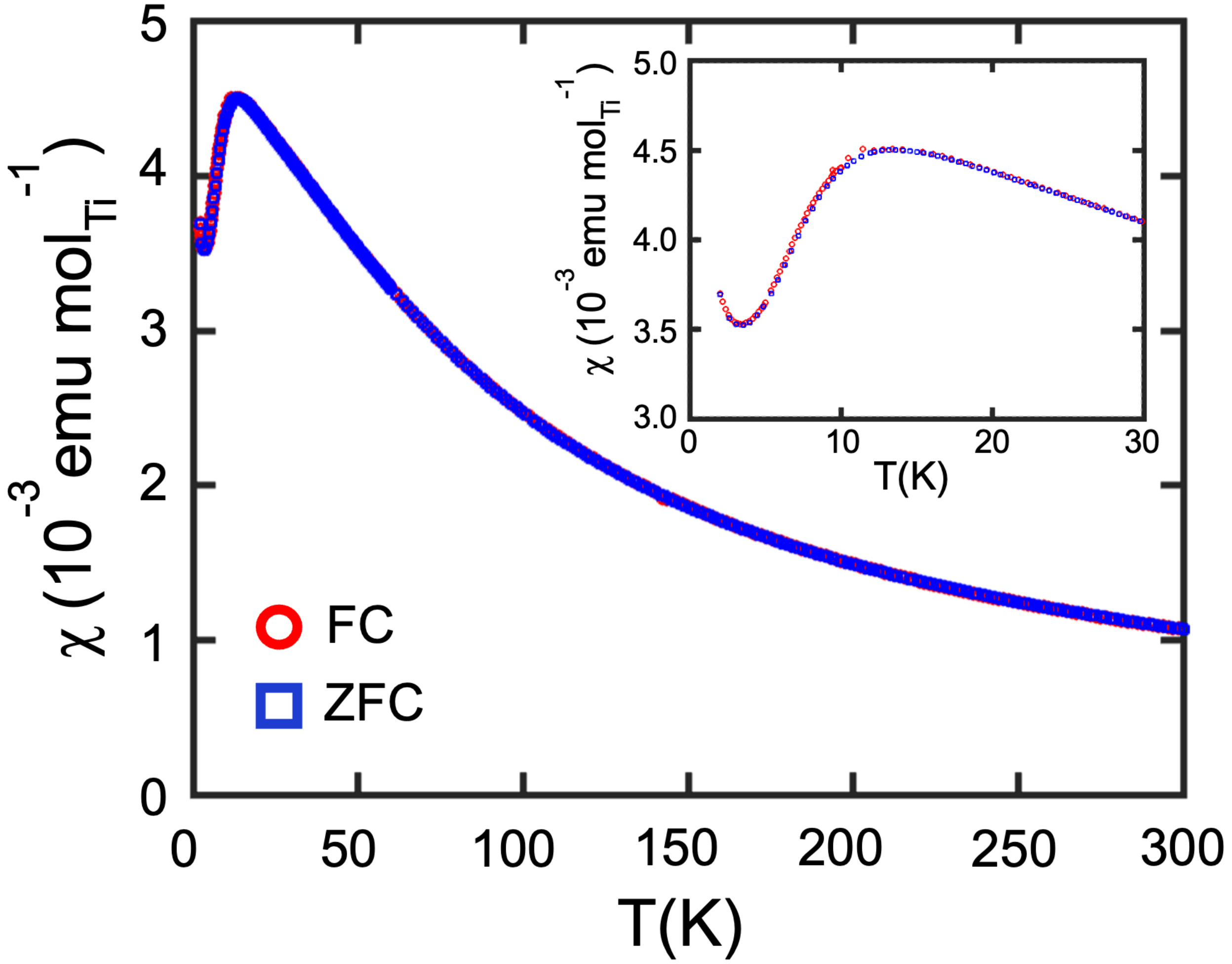

**Extended Data Fig. 2 | Field-Cooled (FC) and Zero-Field-Cooled (ZFC) bulk magnetic susceptibility, $\chi$, as a function of temperature.** $\chi$ was measured under magnetic field of 1 Tesla. In the FC process, the sample was cooled under 1 Tesla. The inset shows the data up to 30 K. The data show no FC-ZFC hysteresis.

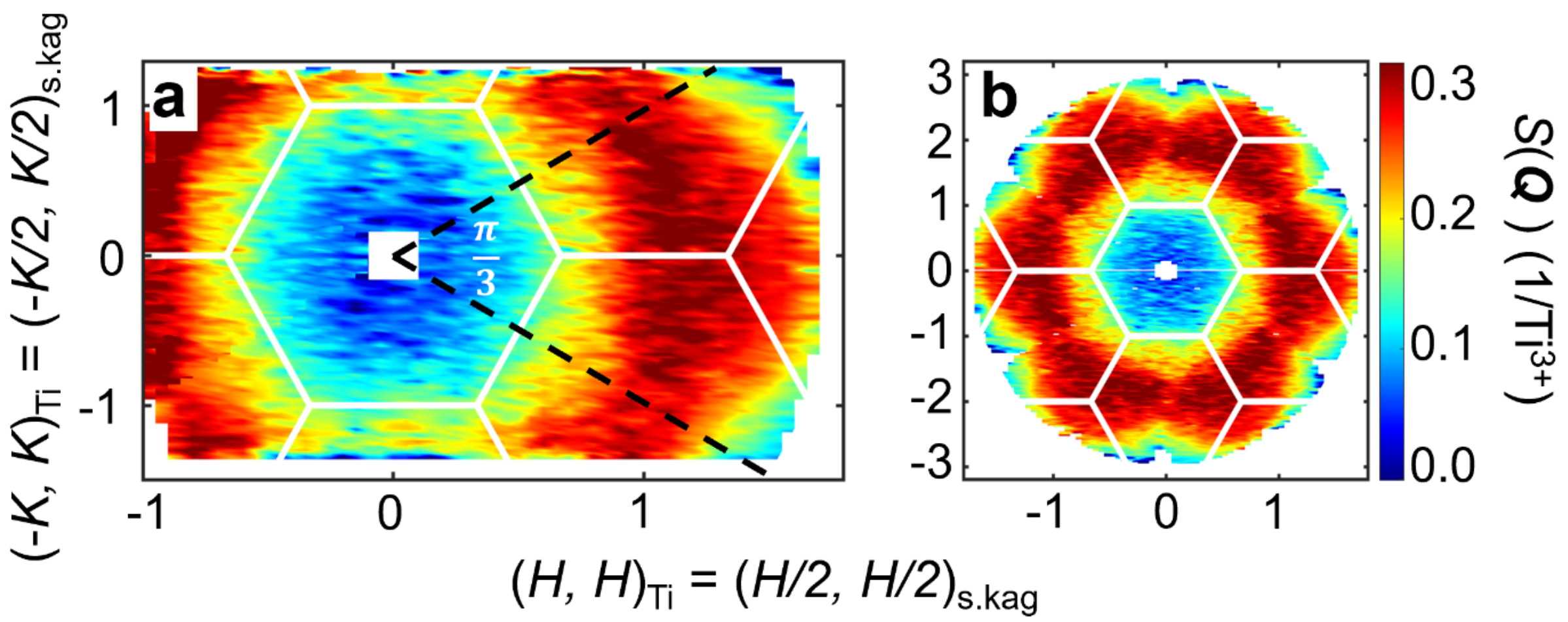

**Extended Data Fig. 3 | Symmetrized Q-dependence of the response function of $Cs_8RbK_3Ti_{12}F_{48}$.** **a**, Contour map of the $\hbar\omega$ -integrated response function, $S(\mathbf{Q}) = \int_{1.9meV}^{2.9meV} S(\mathbf{Q},\hbar\omega)d(\hbar\omega)$ obtained from the experimental data. The black dashed lines represent a $60^0$ sector used to symmetrize the raw data considering the six-fold symmetry of the system. **b**, Contour map of the resulting symmetrized $S(\mathbf{Q})$ data, providing a complete $360^0$ representation.

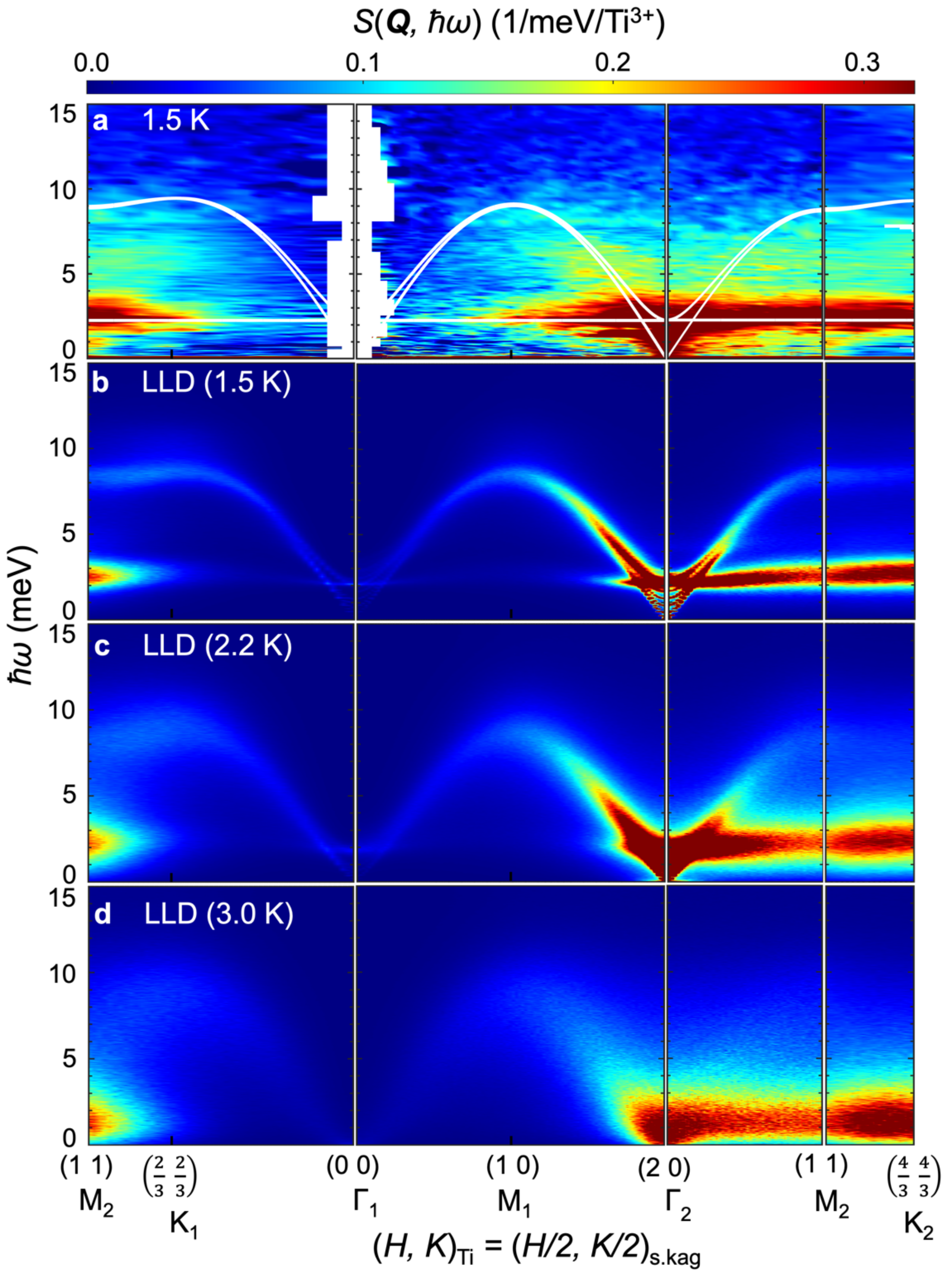

**Extended Data Fig. 4 | Experimental and Landau-Lifshitz dynamics simulated dynamic spin correlation function for $Cs_8RbK_3Ti_{12}F_{48}$. a,** Color contour maps of the normalized experimental dynamic spin correlation function, $S(\mathrm{Q}, \hbar\omega)$, versus momentum (Q) and energy transfer ($\hbar\omega$) along the $\mathrm{M}_2 \to \Gamma_1 \to \Gamma_2 \to \mathrm{M}_2 \to \mathrm{K}_2$ path in the 2D Brillouin zone, taken at 1.5 K. **b-d,** Landau-Lifshitz dynamics simulation results obtained for 1.5 K (**b**), 2.2 K (**c**), and 3 K (**d**) for the spin Hamiltonian of Eq. (1) with $J = 8.7$ meV and $D_{ij} = D\,\hat{z}$ with $D = 0.23$ meV (see main text).

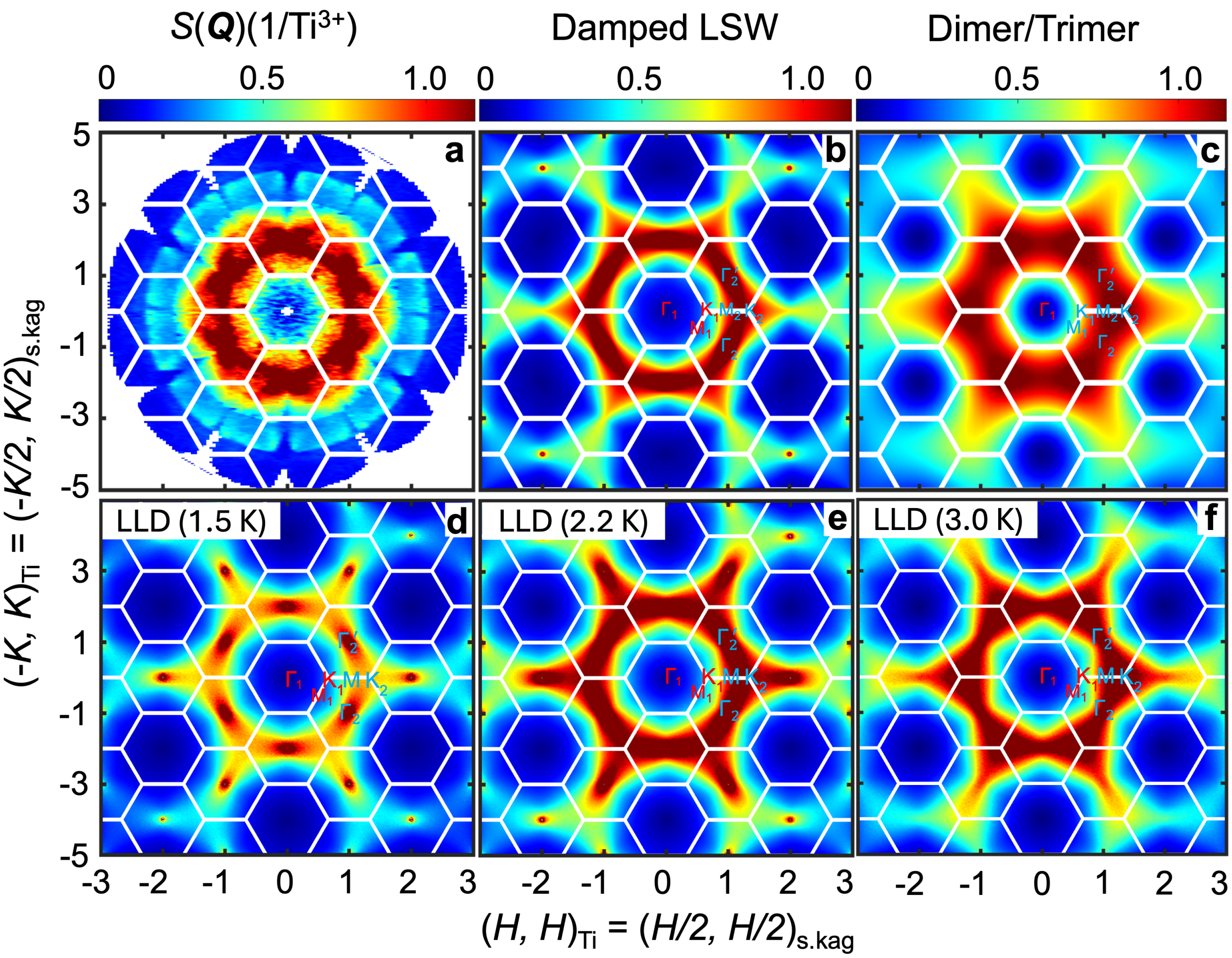

**Extended Data Fig. 5 | Q-dependence of the equal-time response function measured and calculated for $Cs_8RbK_3Ti_{12}F_{48}$. a**, Contour map of the equal-time response function was obtained by performing the following integration, $S(\mathbf{Q}) = \int_{0.3meV}^{13meV} S(\mathbf{Q}, \hbar\omega)\, d(\hbar\omega)$. **b**, Contour map of the corresponding Damped LSW result. **c**, Contour map of the structure factor for decoupled dimers and $120^0$ trimers. **d-f,** Landau-Lifshitz dynamics simulation results obtained for 1.5 K (**d**), 2.2 K (**e**), and 3 K (**f**)

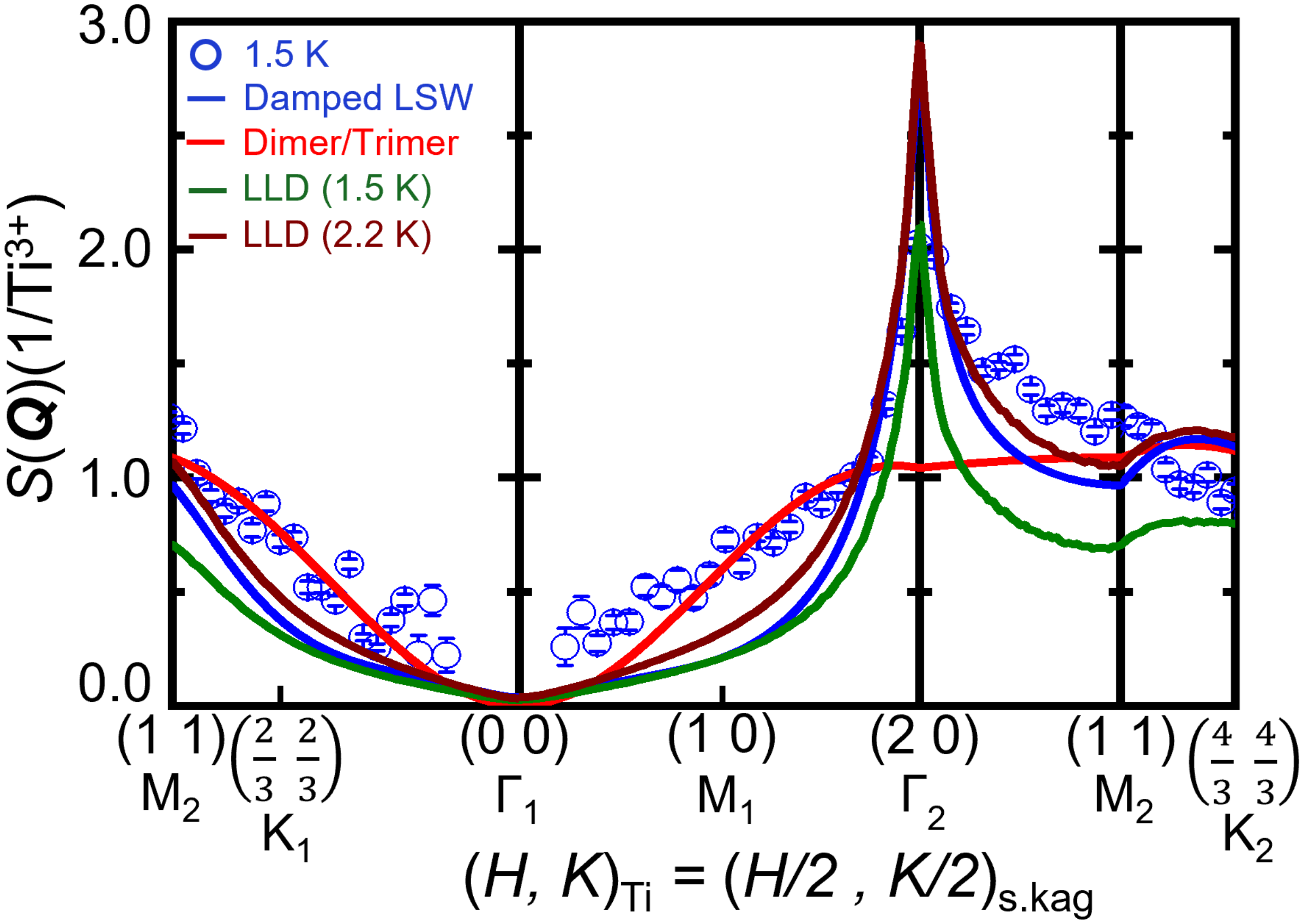

**Extended Data Fig. 6 | Q-dependence of the equal-time response function $\boldsymbol{S}(\boldsymbol{Q})$.** $S(\boldsymbol{Q})$, at 1.5 K, along the $M_2 \to K_1 \to \Gamma_1 \to M_1 \to \Gamma_2 \to M_2 \to K_2$ path through the Brillouin zone. Blue circles are experimental data. Blue and red lines represent the corresponding **Q**-dependence predicted by the damped LSW theory and by decoupled dimers/trimers, respectively, while green and deep violet lines represent the results of LLD simulations at 1.5 K and 2.2 K, respectively.

# Supplementary Information for

# Gapless dispersive continuum in a modulated quantum kagome antiferromagnet

## Contents

## Growth of $Cs_8RbK_3Ti_{12}F_{48}$ single crystals

All the processes for synthesis were carried out under Ar atomosphere in a glove box. Alkali metal halides CsF (99.9%, Sigma-Aldrich), KF (99%, Kojundo Chemical), RbF (99.8%, Sigma-Aldrich), CsCl (99.9%, Sigma-Aldrich), KCl (99.99%, Rare Metallic) and RbCl (Sigma-Aldrich, 99%) were dried at 200 C in Ar flow for 1 h. We synthesized $TiF_3$ through thermal decomposition of $(NH_4)_3TiF_6$ at 600 C for 0.5 h in a cupper container, and $(NH_4)_3TiF_6$ was obtained by heating a mixture of $Ti_2O_3$ and $NH_4HF_2$ (98.5%, Wako Pure Chemicals) in a molar ratio of 1:6.6 in a Teflon container at 220 C for several days in Ar flow, and $Ti_2O_3$ was obtained by heating a stoichiometric mixture of $TiO_2$ (99.99%, Rare Metallic) and Ti (99.9%, Nilaco) at 800C for 1d in an evacuated silica tube. A mixture of CsF, KF, RbF, $TiF_3$, CsCl, KCl and RbCl in a molar ratio of 8:3:1:12:16:6:2 with a total mass of 4 g, was put into a Ni crucible with an inner diameter of 12 mm. The crucible was heated up to 850 C in 2.5 h, was held for 2 h, and was cooled down to 500 C at the rate of 2 C/h. After removing the flux using water, we obtained single crystals, the maximum weight of which was 610 mg.

## Absolute normalization of magnetic neutron scattering data[48] and sum rule

Normalization of the neutron scattering intensity data to absolute units for the scattering cross section can be accomplished through comparison to well-known standards including incoherent elastic scattering from vanadium, sample incoherent elastic scattering, sample elastic nuclear peaks, and sample phonon scattering[48]. We used the sample incoherent elastic scattering as described here with additional details in ref. 48. The neutron scattering intensity measured at the detector can be written as

$$I(\boldsymbol{Q},\hbar\omega) = \int \frac{d^2\sigma}{d\Omega_0 d(\hbar\omega_0)} R(\boldsymbol{Q}_0,\hbar\omega_0,\boldsymbol{Q},\hbar\omega)d^3\boldsymbol{Q}_0 d(\hbar\omega_0)$$

where $R(\boldsymbol{Q}_0,\hbar\omega_0,\boldsymbol{Q},\hbar\omega)$ is the instrument resolution function. And for unpolarized neutrons the coherent magnetic neutron scattering cross-section from a sample with a single species of magnetic atom can be written as

$$\frac{d^2\sigma}{d\Omega d(\hbar\omega)} = N\left(\frac{k_f}{k_i}\right)\left\{\left(\frac{\gamma r_0}{2}\right)gF(\boldsymbol{Q})\right\}^2 e^{-2W(\mathbf{Q})}\sum_{\alpha,\beta}(\delta_{\alpha,\beta} - \hat{\mathbf{Q}}_\alpha\hat{\mathbf{Q}}_\beta)S^{\alpha\beta}(\boldsymbol{Q},\hbar\omega)$$

where $S^{\alpha\beta}(\boldsymbol{Q},\hbar\omega) = \frac{1}{2\pi\hbar}\int dt\, e^{-i\omega t}\sum_l e^{-i\mathbf{Q}\cdot r_l}\langle S_0^\alpha(0)S_l^\beta(t)\rangle$ is the dynamical magnetic structure factor, $N$ is the total number of unit cells, $g = 2$, $F(\mathbf{Q})$ is the magnetic form factor, $\left(\frac{\gamma r_0}{2}\right) = 0.2695\times 10^{-12}$ cm, $k_i$ is the incident neutron wave vector, $k_f$ is the final neutron wave vector, $e^{-2W(\mathbf{Q})}$ is the Debye-Waller factor, $\alpha, \beta$ are the Cartesian coordinates $x$, $y$, and $z$, and $\widehat{\mathbf{Q}}_\alpha, \widehat{\mathbf{Q}}_\beta$ are the projections of the unit wave vector on the Cartesian axes. Once the monitor normalization and the $\left(\frac{k_f}{k_i}\right)$ modification are done, the normalized neutron scattering intensity $\tilde{I}(\boldsymbol{Q}, E)$ can be rewritten approximately as

$$\tilde{I}(\boldsymbol{Q}, \hbar\omega) \approx N \left\{\left(\frac{\gamma r_0}{2}\right) gF(\boldsymbol{Q})\right\}^2 e^{-2W} \tilde{S}(\boldsymbol{Q}, \hbar\omega) R_0$$

where $R_0 = \int R(\boldsymbol{Q}_0, \hbar\omega_0, \boldsymbol{Q}, \hbar\omega) d\boldsymbol{Q}_0 d(\hbar\omega_0)$ is the resolution volume that depends only on the instrument setup, and $\tilde{S}(\boldsymbol{Q}, \hbar\omega)$ is the modified dynamic spin correlation function $\tilde{S}(\boldsymbol{Q}, \hbar\omega) = \sum_{\alpha,\beta}\left(\delta_{\alpha,\beta} - \tilde{Q}_\alpha \tilde{Q}_\beta\right) S^{\alpha\beta}(\boldsymbol{Q}, \hbar\omega)$[48]. Thus, $\tilde{S}(\boldsymbol{Q}, \hbar\omega)$ is given by

$$\tilde{S}(\boldsymbol{Q}, \hbar\omega) = \frac{13.77\ (b^{-1})\ \tilde{I}(\boldsymbol{Q}, \hbar\omega)}{g^2 F(\boldsymbol{Q})^2 e^{-2W} N R_0}$$

where $1\ b = 10^{-24}\ \text{cm}^2$ is the unit for the neutron scattering cross section and $\tilde{S}(\boldsymbol{Q}, \hbar\omega)$ has units of $\text{meV}^{-1}$. Dividing $\tilde{S}(\boldsymbol{Q}, \hbar\omega)$ by the number of magnetic atoms inside one crystal unit cell, we get units of $\frac{1}{meV \cdot magnetic\ atom}$. In the main text, we used $S(\mathbf{Q}, \hbar\omega) = \tilde{S}(\mathbf{Q}, \hbar\omega) F(\mathbf{Q})^2$ to discuss $\boldsymbol{Q}$ dependance or energy dependence of the data in the absolute units. $NR_0$ can be obtained in the absolute unit by considering the incoherent elastic scattering from the sample.

The elastic incoherent scattering cross-section from a sample is given by

$$\left.\frac{d\sigma}{d\Omega}\right|_{inc}^{el} = \frac{N}{4\pi} \sum_j \sigma_j^{inc}\, e^{-2W_j(\mathbf{Q})}$$

where the summation is over all the atoms in the chemical unit cell and $\sigma_j$ is the incoherent neutron scattering cross-section of the $j^{\text{th}}$ atom. Then, the energy-integrated incoherent elastic scattering intensity becomes

$$\int I(\boldsymbol{Q}, \hbar\omega) d(\hbar\omega) = \frac{N}{4\pi} \sum_j \sigma_j^{inc}\, e^{-2W_j(\mathbf{Q})} R_0$$

from which $NR_0$ can be obtained,

$$NR_0 = 4\pi \frac{\int I(\boldsymbol{Q}, \hbar\omega) d(\hbar\omega)}{\sum_j \sigma_j^{inc}\, e^{-2W_j}}$$

Once the magnetic neutron scattering intensity is normalized in the absolute unit, one can check the sum rule. For isotropic magnetic interactions, the sum rule can be the following

$$\frac{\int_{-\infty}^{+\infty}\int_{BZ}\tilde{S}(\boldsymbol{Q},\hbar\omega)d\mathbf{Q}d(\hbar\omega)}{\int_{BZ}\boldsymbol{d}\mathbf{Q}}=\frac{2}{3}s(s+1)$$

where the integration $\int_{BZ}\boldsymbol{dQ}$ is over a Brillouin zone.

**Linear Spin Wave theory with the DFT exchange couplings**

The magnetic ground state and magnetic excitation spectra for the spin Hamiltonian with the Density Functional Theory (DFT) estimated $J$s and antisymmetric Dzyaloshinskii-Moriya (DM) interactions with a fixed ratio of $D_i/J_i = 0.075$ for each bond were determined using the SUNNY package. The Supplementary Fig. 4a shows the ground state that is different from the $q = 0$ ground state for the uniform $J$ shown in Fig. 2c. In the latter, all triangles have a $120^0$ spin arrangement. On the other hand, in the former with the nonuniform $J$s the triangles with uniform $J$s have $120^0$ spin arrangements while the triangles with nonuniform $J$s have spin arrangements different from the $120^0$ configuration, which leads to the enlargement of the magnetic unit cell to be the same as the chemical unit cell and to Bragg peaks and strong low energy gapless magnetic fluctuations at $\mathrm{M}_1 = (10)_{\mathrm{Ti}} = (\frac{1}{2}0)_{\mathrm{s.kag}}$ as shown in Supplementary Fig. 4b. The absence of the low energy gapless magnetic fluctuations at $\mathrm{M}_1 = (10)_{\mathrm{Ti}} = (\frac{1}{2}0)_{\mathrm{s.kag}}$ in the experimental data shown in Fig. 2a indicates a remarkable robustness of the quantum kagome spin liquid state.

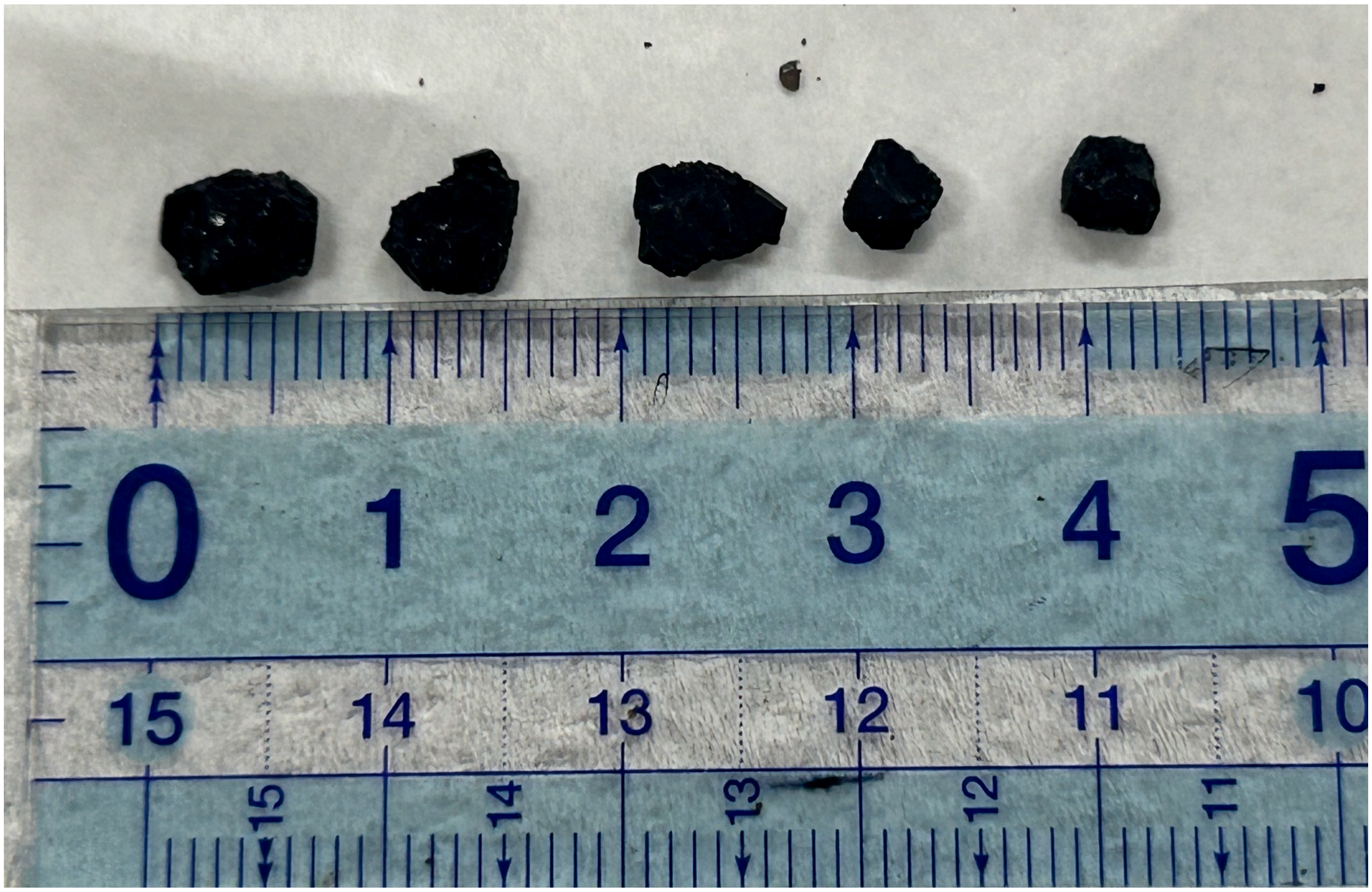

**Supplementary Fig. 1 | Image of single crystals of $Cs_8RbK_3Ti_{12}F_{48}$.** Single crystals of $Cs_8RbK_3Ti_{12}F_{48}$ some of which were used for neutron scattering experiments.

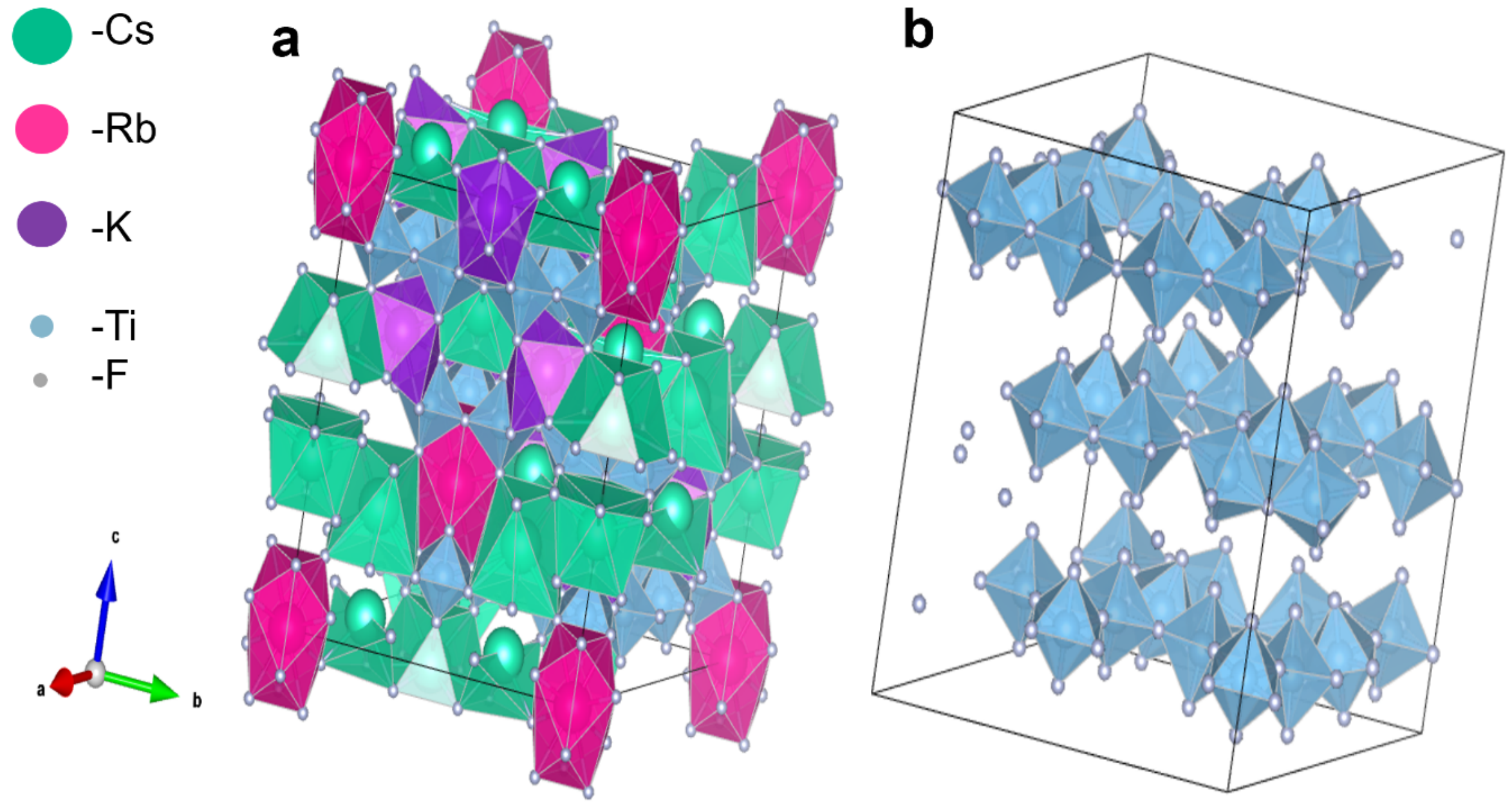

**Supplementary Fig. 2 | Crystal structure of the $Cs_8RbK_3Ti_{12}F_{48}$.** **a**, Polyhedral representation showing all atoms in the crystal lattice. **b**, Zoomed-in view focusing on Ti and F atoms only. The typical distance between the nearest neighboring $Ti^{3+}$ ions in the *ab* kagome is 3.8 Å. The distance between the kagome layers is roughly 6.1 Å.

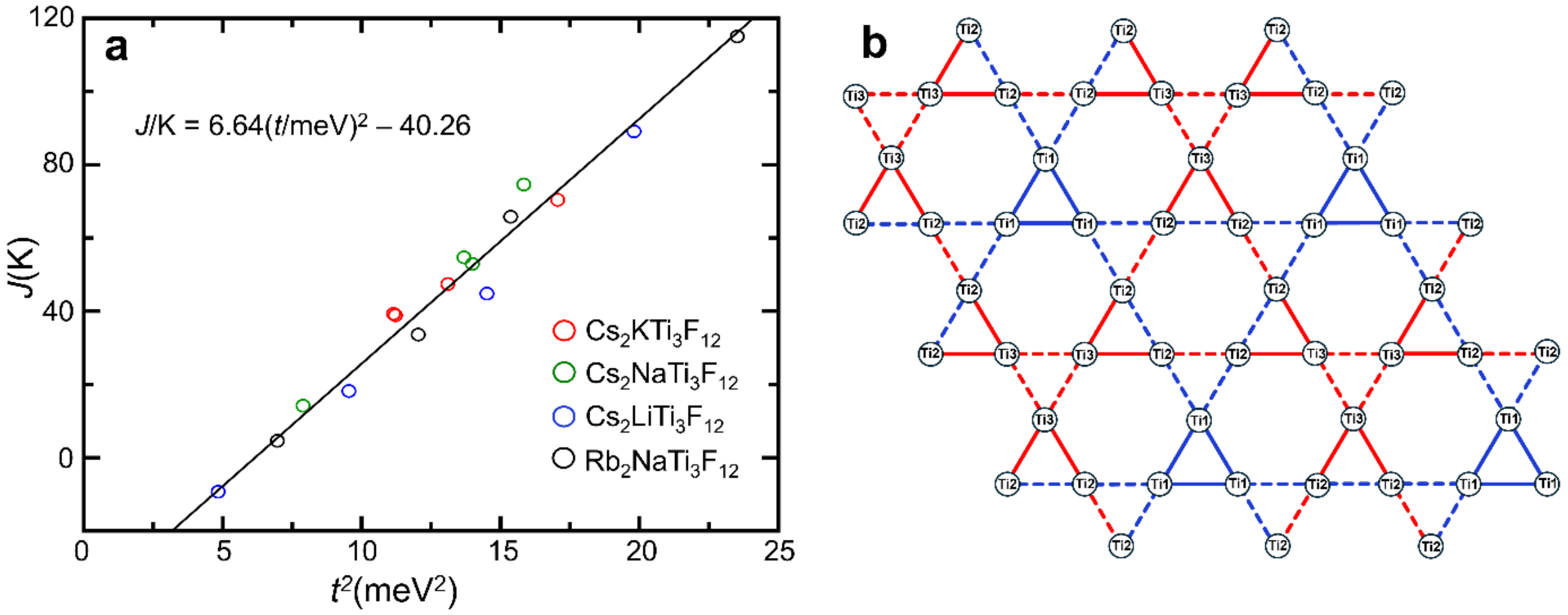

**Supplementary Fig. 3** | **a**, A linear relation between the previously reported near neighbor exchange constants $J$[19,20] and the transfer integral squared $t^2$ for $Cs_2KTi_3F_{12}$ (red), $Cs_2NaTi_3F_{12}$ (green), $Cs_2LiTi_3F_{12}$ (blue), and $Rb_2NaTi_3F_{12}$ (black). **b**, Distorted kagome lattice formed by the three crystallographically distinct $Ti^{3+}$ ions in $Cs_8RbK_3Ti_{12}F_{48}$ that we denote by Ti1, Ti2, and Ti3 as in Supplementary Table 1. The red solid, red dashed, blue solid, and blue dashed lines represent the coupling strengths of $J = (7.4, 6.3, 5.1, 3.9)$ meV, respectively, for the magnetic interactions between the near neighbor $Ti^{3+}$ ions, predicted by Density Functional Theory.

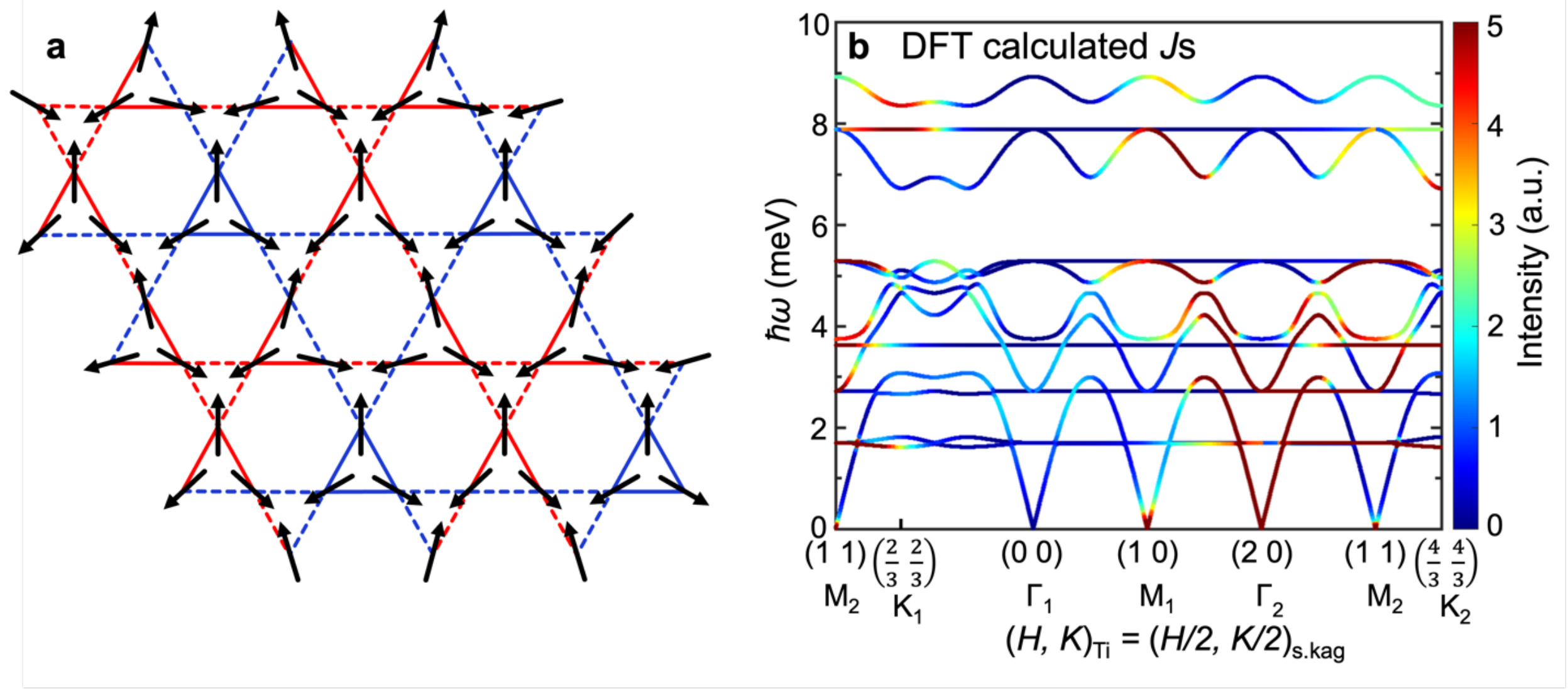

**Supplementary Fig. 4 | a.** Illustration of a distorted kagome lattice formed by the three crystallographically distinct $Ti^{3+}$ ions in $Cs_8RbK_3Ti_{12}F_{48}$, showing the ground-state magnetic structure generated using the SUNNY package. This structure is based on density functional theory (DFT) estimated exchange interactions $J$s (see Supplementary Fig. 3b), with each $J$ assigned a different $D$ at a fixed ratio $D_i/J_i = 0.075$. **b.** Dispersion relations of linear spin waves (LSW) predicted for the DFT-estimated $J$ values and corresponding $D$ values, with intensity color-coded in arbitrary units. Note that there are strong low energy gapless magnetic fluctuations at $\mathrm{M_1} = (10)_{\mathrm{Ti}} = (\frac{1}{2}0)_{\mathrm{s.kag}}$.

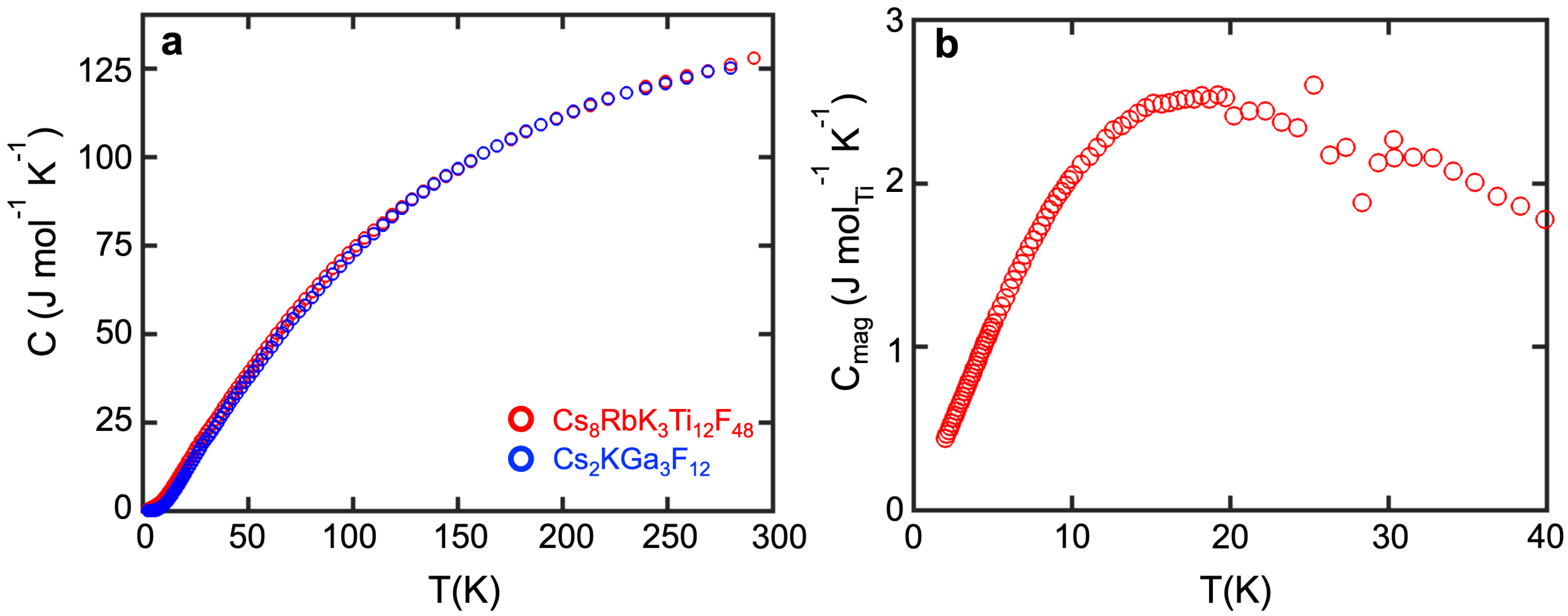

**Supplementary Fig. 5 | Specific heat data. a,** Raw data of specific heat obtained for $Cs_8RbK_3Ti_{12}F_{48}$ (red) and nonmagnetic $Cs_2KGa_3F_{12}$. **b,** Magnetic specific heat for $Cs_8RbK_3Ti_{12}F_{48}$ was obtained by subtracting the nonmagnetic contributions estimated from the $Cs_2KGa_3F_{12}$ data.

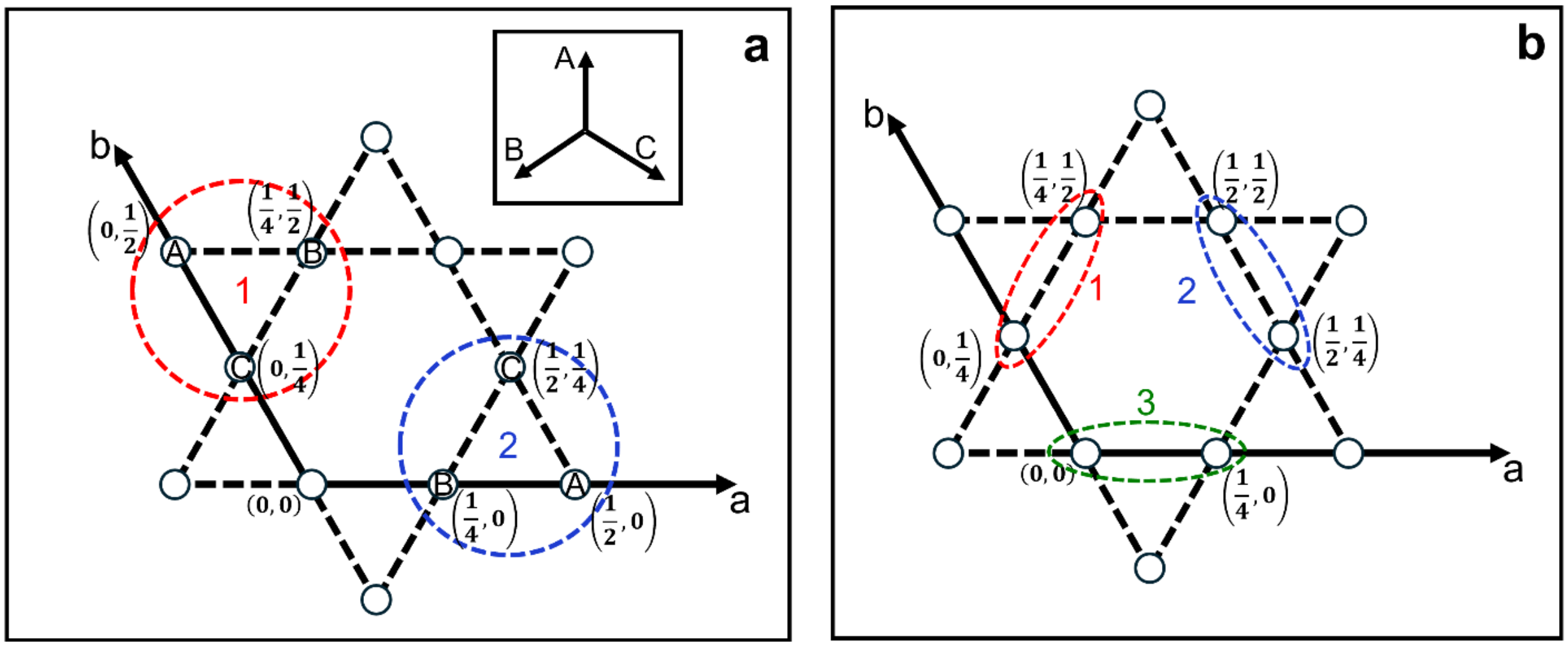

**Supplementary Fig. 6 | Illustration representing decoupled dimers and 120$^0$ decoupled trimers. a**, Trimer model representation of the kagome lattice. The blue dashed circle and red dashed circles represent two trimers used for neutron scattering intensity calculations shown in Extended Data Figures 5 and 6. The three different spins in the 120$^0$ spin structure are labeled A, B, and C as shown in the inset. **b**, Decoupled dimer representation of the kagome lattice. The red, blue, and green dashed ovals represent the three different dimers used for neutron scattering intensity calculations shown in Extended Data Figures 5 and 6.

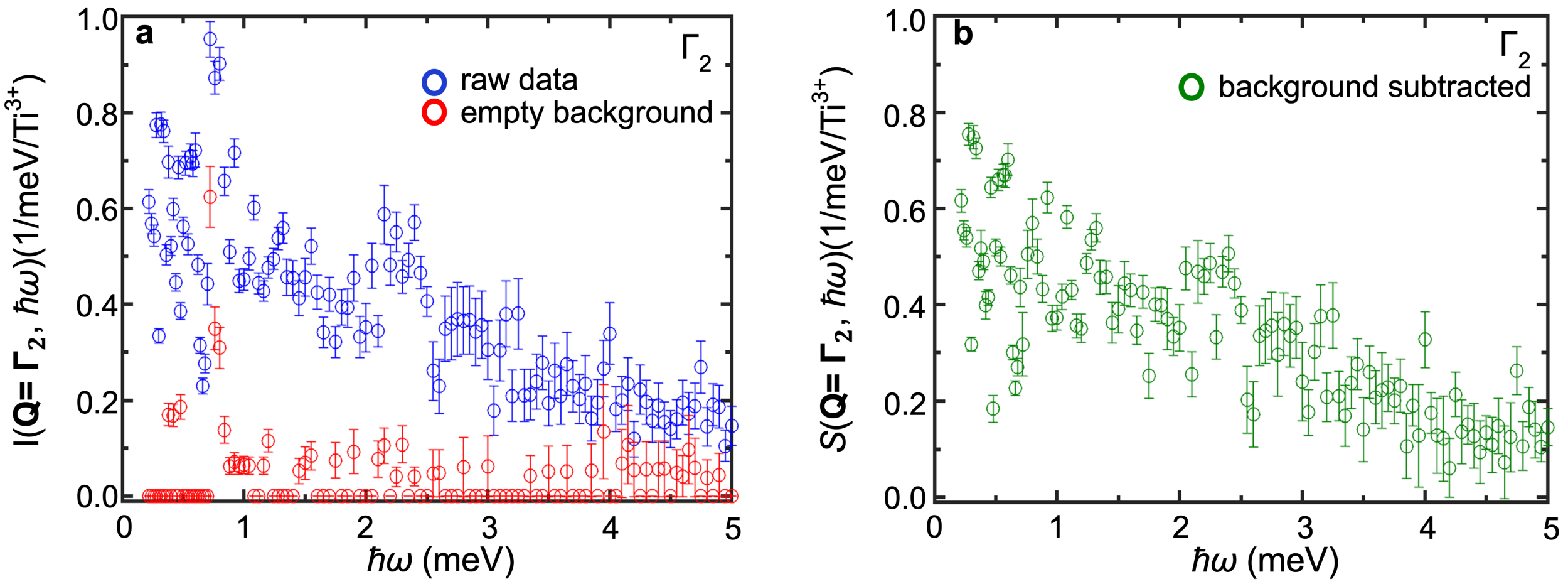

**Supplementary Fig. 7 | $\hbar\omega$- dependence of the dynamic spin correlations at the $\boldsymbol{\Gamma_2}$ point at 1.5 K.** To see if the bump in $S(\boldsymbol{\Gamma_2}, \hbar\omega = 0.8\, meV)$ shown in Fig. 4 d is real or not, here we plotted **a**, the raw data and empty background intensity, $I(\mathbf{Q}, \hbar\omega)$, integrated over $(h \pm 0.05, k \pm 0.05)_{\mathrm{Ti}}$ at $\mathbf{Q} = \boldsymbol{\Gamma_2}$. Note that there is a sharp background at $\hbar\omega = 0.8\, meV$. The blue and red circles represent the raw data and the empty cryostat background, respectively. **b**, The background subtracted magnetic response function, $S(\boldsymbol{\Gamma_2}, \hbar\omega)$, which indicates that the bump centered at $\hbar\omega = 0.8\, meV$ in Fig. 4 d is not real, but it is an artifact due to the high background around $\hbar\omega = 0.8\, meV$ and the resulting poor statistics in $S(\boldsymbol{\Gamma_2}, \hbar\omega)$.

**Supplementary Table 1 | The space group and the structural parameters for $Cs_8RbK_3Ti_{12}F_{48}$.** The parameters were determined using the x-ray data obtained at room temperature over the scattering angle from $\theta_{min}$ to $\theta_{max}$. The numbers in parentheses are standard deviations in the last significant figures. Z is the number of chemical formulas in one chemical unit cell. The total number of Bragg reflections used for the refinement was $N_{total} = 1405$. $R$ is the reliability factor and $\omega R$ is the weighted reliability factor. In general, the refinement is satisfactory when $R < 0.1$ and $\omega R < 0.25$.

| | $Cs_8RbK_3Ti_{12}F_{48}$ |
|---|---|
| Space group | $R3m$ (Hexagonal axes) |
| $a$ (Å) | 15.242 (17) |
| $b$ (Å) | 15.242 (17) |
| $c$ (Å) | 18.476 (19) |
| $\alpha$ (°) | 90.0 (0) |
| $\beta$ (°) | 90.0 (0) |
| $\gamma$ (°) | 120.0 (0) |
| $V$ (Å$^3$) | 3717(9) |
| $Z$ | 3 |
| $\theta_{min}$ (°) | 1.896 |
| $\theta_{max}$ (°) | 30.541 |
| $N_{total}$ | 1405 |
| $R$ | 0.0620 |
| $\omega R$ | 0.1835 |

**Supplementary Table 2 | Atomic coordinates and isotropic displacement parameters for $Cs_8RbK_3Ti_{12}F_{48}$.** They were determined using the x-ray data obtained at room temperature. The numbers in parentheses are standard deviations in the last significant digits.

| Atom | $x$ | $y$ | $z$ | $U$ (Å$^2$) | Occupancy |
|---|---|---|---|---|---|
| Cs1 | 0.16155(6) | 0.32311(13) | 0.7026(4) | 0.0412(5) | 1 |
| Cs2 | 0.000000 | 0.000000 | 0.3759(4) | 0.0383(7) | 1 |
| Cs3 | -0.16775(5) | -0.33550(10) | 0.2740(4) | 0.0363(4) | 1 |
| Cs4 | 0.000000 | 0.000000 | 0.6114(4) | 0.0351(6) | 1 |
| Rb1 | 0.000000 | 0.000000 | 0.000000 | 0.0519(13) | 1 |
| K1 | -0.16311(11) | 0.16311(11) | 0.6530(4) | 0.0191(6) | 1 |
| Ti1 | -0.08336(13) | -0.1667(3) | 0.8319(4) | 0.0202(7) | 1 |
| Ti2 | -0.24995(16) | -0.0001(2) | 0.4812(3) | 0.0184(6) | 1 |
| Ti3 | 0.4993(3) | 0.24965(13) | 0.4981(4) | 0.0179(6) | 1 |
| F1 | -0.1052(6) | -0.3932(6) | 0.5121(8) | 0.045(3) | 1 |
| F2 | 0.0617(5) | 0.1235(10) | 0.8438(14) | 0.065(6) | 1 |
| F3 | -0.3951(6) | -0.1035(6) | 0.5014(9) | 0.051(3) | 1 |
| F4 | -0.0615(5) | -0.1231(10) | 0.1542(11) | 0.053(5) | 1 |
| F5 | -0.2397(5) | -0.4794(11) | 0.3991(9) | 0.065(5) | 1 |
| F6 | 0.1081(4) | 0.2163(8) | 0.4525(7) | 0.0239(19) | 1 |
| F7 | -0.2825(9) | 0.0054(7) | 0.3856(6) | 0.046(2) | 1 |
| F8 | -0.0800(6) | -0.1599(12) | 0.7340(9) | 0.131(14) | 1 |
| F9 | 0.2096(9) | 0.1048(4) | 0.4603(7) | 0.030(2) | 1 |
| F10 | -0.2183(9) | -0.2108(8) | 0.5773(6) | 0.043(2) | 1 |
| F11 | -0.1503(10) | -0.0752(5) | 0.2660(9) | 0.088(8) | 1 |
| F12 | 0.2454(7) | 0.4909(14) | 0.6012(11) | 0.125(13) | 1 |